\documentclass[peerreview]{IEEEtran}
\usepackage[margin=1in]{geometry}
\newgeometry{top=1in,bottom=1in,right=1in,left=1in}

\usepackage[utf8]{inputenc}
\usepackage{amsmath,bm}
\usepackage{amsthm}
\usepackage{amssymb}
\usepackage{mathtools}
\allowdisplaybreaks
\usepackage{graphicx}
\usepackage[dvipsnames]{xcolor}
\usepackage[english]{babel}
\usepackage{tabularx}
\usepackage{multirow}
\usepackage{cite}
\usepackage{enumerate}

\usepackage{caption}
\usepackage{subcaption}

\usepackage{thm-restate}

\usepackage{tikz}\usetikzlibrary{shapes.multipart}
\usetikzlibrary{positioning}
\tikzset{font={\fontsize{9pt}{12}\selectfont}}
\tikzset{>=latex}
\usetikzlibrary{calc}
\usetikzlibrary{arrows}
\usetikzlibrary{external}
\usetikzlibrary{patterns}
\usetikzlibrary{fit}
\usepackage[draft]{changes}

\usetikzlibrary{automata,arrows}

\newcommand{\rcomment}[1]{\textcolor{red}{#1}}

\DeclareMathOperator{\ir}{Irr}

\DeclareMathOperator{\mds}{MDS}

\newtheorem{theorem}{Theorem$\!$}
\newtheorem{example}[theorem]{Example$\!$}
\newtheorem{lemma}[theorem]{Lemma$\!$}

\newtheorem{construction}{Construction$\!$}

\newtheorem{definition}[theorem]{Definition$\!$}
\usepackage{graphicx}

\usepackage{xcolor}

\newcommand{\bcomment}[1]{{\leavevmode\color{blue}#1}}

\newcommand{\mA}{\mathsf{A}}
\newcommand{\mC}{\mathsf{C}}
\newcommand{\mG}{\mathsf{G}}
\newcommand{\mT}{\mathsf{T}}

\newcommand{\cB}{\mathcal{B}}
\newcommand{\cC}{\mathcal{C}}
\newcommand{\cD}{\mathcal{D}}
\newcommand{\cE}{\mathcal{E}}

\newcommand{\cL}{\mathcal{L}}

\newcommand{\cR}{\mathcal{R}}

\newcommand{\cT}{\mathcal{T}}
\newcommand{\cU}{\mathcal{U}}

\newcommand{\bbF}{\mathbb{F}}

\newcommand{\mybold}[1]{\bm{#1}}

\newcommand{\bb}{{\mybold{b}}}
\newcommand{\bc}{{\mybold{c}}}

\newcommand{\br}{{\mybold{r}}}
\newcommand{\bs}{{\mybold{s}}}
\newcommand{\bt}{{\mybold{t}}}
\newcommand{\bu}{{\mybold{u}}}
\newcommand{\bv}{{\mybold{v}}}
\newcommand{\bw}{{\mybold{w}}}
\newcommand{\bx}{{\mybold{x}}}
\newcommand{\by}{{\mybold{y}}}
\newcommand{\bz}{{\mybold{z}}}

\newcommand{\balpha}	{\mybold{\alpha}}
\newcommand{\bbeta}		{\mybold{\beta}}

\newcommand{\bsigma}	{\mybold{\sigma}}

\DeclareMathOperator{\nmod}{mod}
\newcommand{\shiftedMod}{\,\mathbin{\nmod^+}\,}

\newcommand{\nblk}{{N_B}}  % The number of message blocks in Section V

\newcommand{\ddes}[1]{\mathcal D^{#1}}  % duplication-substitution
\newcommand{\ddeo}{\mathcal D}  % duplication channel

\newcommand{\ddesded}[2]{\mathcal D^{#1}_{#2}} % duplication-substitution-deduplication-substitution channel

\newcommand{\codeA}{\cC^A}
\newcommand{\codeB}{\cC^B}
\newcommand{\codeJain}{\cC^D}
\newcommand{\dec}{\operatorname{Dec}}

\begin{document}

\IEEEoverridecommandlockouts
\title{Low-redundancy codes for correcting multiple short-duplication and edit errors}

\author{
  \IEEEauthorblockN{Yuanyuan Tang%
                    \IEEEauthorrefmark{1},
                     Shuche Wang%
                     \IEEEauthorrefmark{2},
                     Hao Lou%
                    \IEEEauthorrefmark{1},
                     Ryan Gabrys%
                     \IEEEauthorrefmark{3},
                     and Farzad Farnoud%
                     \IEEEauthorrefmark{1}
                     }  \\
  \IEEEauthorblockA{\IEEEauthorrefmark{1}
                    Electrical \& Computer Engineering,
                    University of Virginia, U.S.A.,\\     \texttt{\{yt5tz,hl2nu,farzad\}@virginia.edu}} \\
 \IEEEauthorblockA{\IEEEauthorrefmark{2}
                    Institute of Operations Research and Analytics,
                    National University of Singapore,
                    \texttt{shuche.wang@u.nus.edu}}    \\
  %\IEEEauthorblockA{\IEEEauthorrefmark{2}
  %the California Institute for Telecommunications and Information Technology, UCSD, U.S.A., \texttt{ rgabrys@eng.ucsd.edu}                    }\\
    \IEEEauthorblockA{\IEEEauthorrefmark{3}
  Calit2, University of California-San Diego, U.S.A., \texttt{rgabrys@ucsd.edu} }
  \thanks{This work was supported in part by NSF grants under grant CIF 1816409 and~CIF~1755773.}
  \thanks{ This paper was presented in part at the IEEE ISIT2021 \cite{tang2021error_atmost_p} and  IEEE ISIT2022~\cite{Tang2022Correcting}.}
}

\maketitle

\begin{abstract}
Due to its  higher data
density, longevity, energy efficiency, and ease of generating copies, DNA is considered a promising storage technology for satisfying future needs. However, a diverse set of errors including deletions, insertions, duplications, and substitutions may arise in DNA at different stages of data storage and retrieval. 
The current paper constructs error-correcting codes for simultaneously correcting short (tandem) duplications and at most $p$ edits, where a short duplication generates a copy of a substring with length $\leq 3$ and inserts the copy following the original substring, and an edit is a substitution, deletion, or insertion.
Compared to the state-of-the-art codes for duplications only, the proposed codes correct up to $p$ edits (in addition to duplications) at the additional cost of roughly $8p(\log_q n) (1+o(1))$ symbols of redundancy, thus achieving the same asymptotic rate, where $q\geq 4$ is the alphabet size and $p$ is a constant. Furthermore, the time complexities of both the encoding and decoding processes are polynomial when $p$ is a constant with respect to the code length.

\end{abstract}

\section{Introduction}

With recent advances in sequencing and synthesis,deoxyribonucleic acid (DNA) is considered a promising candidate for satisfying future data storage needs~\cite{yazdi2015dna,jain2017noise}. In particular,  experiments in~\cite{yazdi2015dna,yazdi2015rewritable, erlich2017dna, blawat2016forward, organick2018random, lee2019terminator} demonstrate that data can be stored on and subsequently retrieved from DNA. Compared to traditional data storage media,  DNA has the advantages of higher data density, longevity, energy efficiency, and ease of generating copies~\cite{yazdi2015dna,lee2019terminator}. 
However, a diverse set of errors may occur at different stages of the data storage and retrieval processes, such as deletions, insertions, duplications, and substitutions.  Many recent works, such as~\cite{jain2017duplication,shipman2017, kovavcevic2018asymptotically, yehezkeally2019reconstruction, tang2020single,lenz2019coding, cai2019optimal, elishco2019bounds,lenz2020coding, lee2019terminator, gabrys2020mass,jain2020coding,kiah2020coding,yehezkeally2020uncertainty, nguyen2020constrained, sima2020robust, chee2020efficient,Tang2021CorrectingDeletion}, have been devoted to protecting the data against these errors.
The current paper constructs error-correcting codes for duplication and edit errors, where an edit error is an insertion, deletion, or substitution.

A \emph{(tandem) duplication} in a DNA sequence generates a copy of a substring and then inserts it directly following the original substring~\cite{jain2017duplication}, where the duplication length is the length of the copy. For example, given $\mA\mC\mT\mG$, a tandem duplication may generate $ \mA\mC\mT\underline{\mC\mT}\mG$, 
where $\mC\mT\underline{\mC\mT}$ is a \emph{(tandem) repeat} of length 4 (i.e., twice the length of the duplication). Bounded-length duplications are those whose length is at most a given constant. In particular, we refer to duplications of length at most 3 as \emph{short duplications}. Correcting fixed-length duplications~\cite{jain2017duplication,yehezkeally2019reconstruction,kovavcevic2018asymptotically, tang2020single,Tang2021Noisy} and bounded-length duplications~\cite{jain2017duplication, jain2017capacity, kovavcevic2019codes, chee2019deciding, tang2020error,chee2020efficient} have been both studied recently. In particular, the code in~\cite{jain2017duplication}, which has a polynomial-time encoder, provides the highest known asymptotic rate  for correcting any number of short duplications. For an alphabet of size 4, corresponding to DNA data storage, this rate is  $ \log 2.6590$ and as the alphabet size $q$ increases, the rate is  approximately $\log(q-1)$~\cite{chee2020efficient}.

For channels with both duplication and substitution errors, \emph{restricted} substitutions~\cite{tang2020single, Tang2021Noisy}, which  occur only in duplicated copies, and \emph{unrestricted} substitutions~\cite{tang2020single,tang2019error,tang2020error, Tang2021ECC_Edit}, which may occur anywhere, have been studied.
The closest work to the current paper,~\cite{Tang2021ECC_Edit}, constructed error-correcting codes for short duplications and at most one (unrestricted) edit. However, compared to the codes in~\cite{jain2017duplication} for only duplications, the codes in~\cite{Tang2021ECC_Edit} incur an %non-negligible
asymptotic rate loss when $q=4$ in order to correct the additional edit. The current paper provides codes for correcting any number of short duplications and at most $p$ (unrestricted) edits with no asymptotic rate penalty, where $p$ and the alphabet size $q$ are constants.

One of the challenging aspects of correcting multiple types of errors, even when optimal codes for individual error types exist, is that codes for each type may utilize incompatible strategies. In particular, correcting duplications relies on constrained codes (local constraints) while edits are corrected using error-correcting codes with codewords that satisfy certain global constraints.  Combining these strategies is not straightforward as encoding one set of constraints may violate the other, or alter how errors affect the data. Our strategy, which can be viewed as modified concatenation described in~\cite{marcusintroduction2001}, is to first encode user data as a constrained sequence $\bx$ that is irreducible, i.e., does not contain any repeats of length $\leq 6$. Then using \emph{syndrome compression}, we compute and append to $\bx$ a ``parity'' sequence $\br$ to help correct errors that occur in $\bx$. Syndrome compression has recently been used to provide explicit constructions for correcting a wide variety of errors with redundancy as low as roughly twice the Gilbert-Varshamov bound \cite{sima2020syndrome,sima2020optimal,sima2020optimalCodes,sima2020optimal_Systematic}. Another challenge arises from the interaction between the errors. When both short duplications and edits are present, a single edited symbol may be duplicated many times and affect an unbounded segment. However, when the input is an irreducible sequences, after removing all tandem copies with length  $\leq 3$ from the output, the effects of short duplications and at most $p$ edits can be localized in at most $p$ substrings, each with length $\leq 17$~\cite{Tang2021ECC_Edit}. Using the structure of these localized alterations, we describe the set of strings that can be confused with $\bx$ and bound its size, allowing us to leverage syndrome compression to reduce redundancy.
A third challenge is ensuring that the appended vector $\br$ is itself protected against errors and can be decoded correctly. We do this by introducing a higher-redundancy MDS-based code over irreducible sequences. After decoding the appended vector, we use it to recover the data by eliminating incorrect confusable inputs.
Compared to the explicit code for short duplications only~\cite{jain2017duplication}, the proposed code corrects $\leq p$ edits in addition to the duplications at the extra cost of roughly $8p(\log_q n)(1+o(1))$ symbols of redundancy for $q\geq 4$, and achieves the same asymptotic code rate. We note that the state-of-the-art redundancy for correcting $p$ edits is no less than $4p\log_q n(1+o(1))$~\cite{sima2020optimalCodes}. Time complexities of both the encoding and decoding processes are polynomial when $p$ is a constant.

For simplicity, we first consider the channel with short duplications and \textit{substitutions} only and construct codes for it. Then, in Subsection~\ref{ssc:generalize}, we show that the same codes can correct short duplications and \textit{edit} errors. We note that short duplications and edits may occur in any order. Henceforth, the term duplication refers to short duplications only.

The paper is organized as follows. Section~\ref{sec:Not_Pre}
presents the notation and preliminaries. %, including a brief review of the syndrome compression technique.
In Section~\ref{Sec:channel_model}, we derive an upper bound on the size of the confusable set for an irreducible string,
%where the confusable set consists of all confusable repeat-free strings of the input
which is a key step of the syndrome compression technique used to construct our
error-correcting codes. Then, Section~\ref{sec:ECC} presents the code construction as well as a discussion of the redundancy and the encoding/decoding complexities, under the assumption that the syndrome information can be recovered correctly by an auxiliary error-correcting code, which is described in Section~\ref{Sec:Aux_ECC}. Finally, Section~\ref{sec:conclusion} concludes the main results.

 %%%%%%%%%%%%%%%%%%%%%%%%%%%%%%%%%%%%%%%%%%%%%%%%%

 \section{Notation and Preliminaries} \label{sec:Not_Pre}

Let $\Sigma_q=\{0,1,2,\cdots, q-1\}$ represent a finite alphabet of size $q$ and $\Sigma_{q}^{n}$ the set of all strings of length $n$ over $\Sigma_q$. Furthermore, let $\Sigma_{q}^{*}$ be the set of all finite strings over $\Sigma_q$, including the empty string $\Lambda$. %, which includes %Then we have $\Sigma_{q}^{*}=\bigcup_{n=0}^{\infty} \Sigma_{q}^{n}$.
 %the empty string $\Lambda$%is also a member of the set $\Sigma_{q}^{*}$, i.e.,
%satisfies $\Lambda \in \Sigma^{*}_q$
Given two integers $a,b$ with $a\leq b$, the set $\{a,a+1, \dotsc, b\}$ is shown as $[a,b]$. We simplify $[1,b]$ as $[b]$. For an integer $a\geq 1$, we define $b \shiftedMod a$ as the integer in $[a]$ whose remainder when divided by $a$ is the same as that of $b$. Unless otherwise stated, logarithms are to the base 2.%$\log(\cdot)$ represents $\log_2 (\cdot)$.
%\rcomment{Yuanyuan: I still suggest using $\log$ representing $\log_2$}

We use bold symbols to denote strings over $\Sigma_{q}$, i.e., $\bx, \by_j\in \Sigma_q^*$. The entries of a string are represented by plain typeface, e.g., the $i$th elements of $\bx, \by_j\in \Sigma_q^*$ are $x_i,y_{ji}\in \Sigma_q$, respectively. For two strings $\bx, \by\in \Sigma_q^*$, let $\bx\by$ denote their concatenation. Given four strings $\bx, \bu,\bv,\bw \in \Sigma^{*}_q$, if  $\bx=\bu\bv\bw$, then $\bv$ is called a substring of $\bx$. %For a string $\bx$, two substrings $\bu$ and $\bv$ are said to \emph{overlap} if we can write $\bx=\ba\bb\bc\bd\be$, where $\bc$ is nonempty, $\bu = \bb\bc$, and $\bv=\bc\bd$.
Furthermore, we let $|\bx|$ represent the length of a string $\bx\in \Sigma_q^n$, and let  $\Vert S \Vert$ denote the size (the number of elements) of a set $S$.

A \emph{(tandem) duplication} of length $k$ is the operation of generating a copy of a substring and inserting it directly following the substring, where $k$ is the length of the copy. For example, for $\bx=\bu \bv \bw$ with $|\bv|=k$, a (tandem) duplication may generate $\bu \bv \bv\bw$, where $\bv\bv$ is called a \emph{(tandem) repeat} with length $2k$. A duplication of length at most $3$ is %denoted as a $\leq\!\! 3$-TD, which we
called \emph{a short duplication}. Unless otherwise stated, the short duplications are simply called duplications in the rest of the paper.
For example, given $\bx=213012\in \Sigma_{4}^*$, a sequence of duplications may produce
\begin{equation} \label{eq:exam_upperk_TD}
\begin{split}
        \bx&=213012\to 213\underline{213}012\to
        2132130\underline{30}12\\& \to 2132\underline{2}1303012=\bx',
\end{split}
\end{equation}
where the duplicated copies %with length upper bounded by $3$
are marked with underlines. We call $\bx'$ a \emph{descendant} of $\bx$, i.e., a string generated from $\bx$ by a sequence of  duplications. Furthermore, for a string $\bx\in \Sigma_q^*$, %let $D_{\leq k}^{\alpha}(\bx)$ be the set that consists of all the descendants generated from $\bx$ by $\alpha$ $\leq\!\! k$-TDs, and similarly
let $\ddeo(\bx)\subseteq \Sigma_q^*$ be the set of all descendants generated from $\bx$ by an arbitrary number of duplications. Note that, unless $\bx=\Lambda$, $\cD(\bx)$ is an infinite set.

A \emph{deduplication} of length $k$ replaces a repeat $\bv\bv$ by $\bv$ with $|\bv|=k$. In the rest of the paper, unless otherwise stated, dedulications are assumed to be of length at most 3. For example, the string $\bx$ in~\eqref{eq:exam_upperk_TD} can be recovered from $\bx'$ by three  deduplications.

The set of \emph{irreducible strings} of length $n$ over $\Sigma_q$, denoted $\ir_q(n)$, consists of strings without repeats $\bv\bv$, where $|\bv|\leq 3$. Furthermore, $\ir_q(*)$ represents all irreducible strings of finite length over $\Sigma_q$. The \emph{duplication root} of $\bx'$ is an irreducible string $\bx$ such that $\bx'$ is a descendant of $\bx$. Equivalently, $\bx$ can be obtained from $\bx'$ by performing all possible deduplications. Any string $\bx'$ has one and only one duplication root~\cite{jain2017duplication}\footnote{Note that this statement only applies to duplications of length at most 3. For duplications of length at most 4, the root is not necessarily unique.}, denoted $R(\bx')$. %The set of duplication roots of $\bx'$ is denoted $R(\bx')$, i.e., $R(\bx')=\{\bx\in \ir_q(*)\mid \bx'\in \ddeo(\bx)\}$. %Note that $R_{\leq k}(\bx')\subseteq \ir_{\leq k}(*)$.
 The uniqueness of the root  implies that if $\bx''$ is a descendant of $\bx'$, we have $R(\bx')= R(\bx'')$. For a set $S$ of strings, we define $R(S)=\{R(\bs):\bs\in S\}$ as the set of the duplication roots of the elements of $S$.

Besides duplications, we  also consider substitution errors, where each substitution replaces a symbol by another one from the same alphabet. Continuing the example in~\eqref{eq:exam_upperk_TD}, two substitutions and two duplications applied to $\bx'$ may produce
 \begin{equation*}
    \begin{split}\label{eq:Example_upperk_TD_sub}
         \bx'&=213221303012 \to 2132\rcomment{1}1303012 \\&\to 2132\rcomment{1} \underline{32\rcomment{1}}1303012
         \to 2132\rcomment{1} 32\rcomment{1} 13\rcomment{2}3012 \\&\to 2132\rcomment{1} 32\rcomment{1} 13\rcomment{2}3 \underline{3\rcomment{2}3}012 =\bx'',
    \end{split}
\end{equation*}
where the substituted symbols are marked in red. Let $\ddes{\leq p}(\bx)\subseteq \Sigma_q^*$ represent the set of strings derived from $\bx$ by an arbitrary number of duplications and at most $p$ substitutions.
In the example above, we have $\bx'' \in \ddes{\leq 2}(\bx)$. Note that the alphabet over which $\ddes{\leq p}(\bx)$ is defined affects its contents. For example, for $\bx=012$, $\ddes{\le 1}(\bx)$ contains $013$ if the alphabet is $\Sigma_4$ but not if the alphabet is $\Sigma_3$. Unless $\bx=\Lambda$, $\ddes{\le p}(\bx)$ is infinite. %Furthermore, let $D_{\leq k}^{*,\leq p}(\bx)$ represent the set of strings generated by an arbitrary number of $\leq\!\!k$-TDs and at most $p$ substitutions.

%AAAAAAAAAAAAAAAAAAAAAAAAAAAAAAAAAAAAAAAAAAAAAAAAAAAAAAAAAAAAAAAAAAAAAAAAAA

%\rcomment{Yuanyuan: not update!}

%For simplicity,  when $k=3$, we drop the $\le\!\!3$ subscript and write $D^*(\cdot), R(\cdot), \ir(\cdot), D^{\alpha, p}(\bx)$, and $D^{*,\leq p}(\bx)$. %In the rest of the paper, unless otherwise stated, duplications are assumed to be $\leq \!\!3$-TDs, deduplications remove copies with length bounded by $3$, and irreducible strings represent $\le\!\!3$-irreducible strings.

We define a \emph{substring edit} in a string $\bx\in\Sigma_q^*$ as the operation of replacing a substring $\bu$ with a string $\bv$, where  at least one of $\bu,\bv$ is nonempty. The length of the  substring edit is $\max\{|\bu|,|\bv|\}$. An \emph{$L$-substring edit} is one whose length is at most $L$. For example, given $\bx=0123456$, a $4$-substring edit can generate the sequence $\by=0\underline{789}56$ or the sequence $\bz = 01\underline{89}23456$, where the inserted strings are underlined. %\bcomment{And another $4$-substring edit may also generate $\bz=021312313$ from $\bx$ by replacing $\bx_{[7,9]}=3120$ with $\bz_{[7]}=3$}.
Furthermore, a \emph{burst deletion} in $\bx\in \Sigma_{q}^*$ is defined as removing a substring $\bv$ of $\bx$, where $|\bv|$ is the length of the burst deletion. %A \emph{$\leq\!\! L$-burst deletion} has length at most $L$. For example, given $\bx=021 3120 2013$, a $\leq\!\! 4$-burst deletion may generate $\by=021 2013$ by removing the substring $\bx_{[4,7]}=3120$.

Given a sequence $\bx \in \Sigma_q^n$, we define the binary matrix $\cU(\bx)$ of $\bx$ with dimensions $\lceil\log q\rceil\times n$ as
\begin{align}\label{eq:binary_matrix}
\begin{bmatrix}
u_{1,1} & u_{1,2}& \dotsm & u_{1,n}  \\
u_{2,1} & u_{2,2} & \dotsm & u_{2,n}  \\
\vdots & \vdots & \ddots & \vdots  \\
u_{\lceil\log q\rceil,1} & u_{\lceil\log q\rceil,2} & \dotsm & u_{\lceil\log q\rceil,n}  \\
\end{bmatrix},
\end{align}
\iffalse % AAAAAAAAAAAAAAAAAAAAAAAAAAAAAAAAAAAAAAAAAAAAAAAAAAAAAAAAAAAA
\begin{equation}\label{eq:binary_matrix}
\begin{bmatrix}
u_{1,1} & u_{1,2}& \dotsm & u_{1,n}  \\
u_{2,1} & u_{2,2} & \dotsm & u_{2,n}  \\
\vdots & \vdots & \ddots & \vdots  \\
u_{\lceil\log q\rceil,1} & u_{\lceil\log q\rceil,2} & \dotsm & u_{\lceil\log q\rceil,n}  \\
\end{bmatrix} \in \{0,1\}^{\lceil\log\bcomment{_2} q\rceil\times n},
\end{equation}
\fi %AAAAAAAAAAAAAAAAAAAAAAAAAAAAAAAAAAAAAAAAAAAAAAAAAAAAAAAAAA
where the $j$th column of $\cU(\bx)$ is the binary representation of the $j$th symbol of $\bx$ for $j\in[n]$. The $i$th row of  $\cU(\bx)$ is denoted as $\cU_i(\bx)$ for $i\in \lceil\log q\rceil$.

%%%%%%%%%%%%%%%%%%%%%%%%%%%%%%%%%%%%%%%%%%%%%%%%%
%%%%%%%%%%% The code consisting of irreducible strings of a given length has the same asymptotic rate as the existing code.

%\bcomment{
The redundancy of a code $\cC\subseteq \Sigma^{n}_q$ of length $n$ is defined as $n-\log_q \Vert \cC\Vert$ symbols, and its rate as $\frac{1}{n}\log \Vert \cC\Vert$ bits per symbol. Asymptotic rate is the limit superior of the rate as the length $n$ grows.%, the code rate and the asymptotic code rate of the code $\cC$ is defined as $R(\codeB_n)=\frac{1}{n}\log \Vert \cC \Vert$ and $R(\cC)=\lim_{n\to \infty}\frac{1}{n}\log \Vert \cC \Vert$, respectively. Furthermore, the redundancy of the code is $r(\codeB_n)=1-R(\codeB_n)$. Then we have the following lemma.

%\rcomment{Yuanyuan: will we modify the definition of code rate since the code rate in Section V is computed over $\log_2$ instead of $\log_q$?}

In order to construct error-correcting codes by applying the syndrome compression technique~\cite{sima2020syndrome}, we first introduce some auxiliary definitions and a theorem.

%\rcomment{Yuanyuan: the explanation and Definition~1 above are easy to misunderstand. I misunderstand it at the beginning since I thought $\cC$ should be able to correct errors. In Ryan's paper, the defined confusable set $B(\bx)$ does not overlap with a code $C$. It may be better to replace the name code as some set $S=\ir(n)\subseteq \Sigma_{q}^n$. Finally, without any analysis, is it good to directly say that we construct codes over irreducible strings.}

%\bcomment{
Suppose $q\geq 3$ is a constant. We start with the definition of confusable sets for a given channel and a given set of strings $S\subseteq \Sigma_q^n$. In our application, $S$ is the set of irreducible strings, upon which the proposed codes will be constructed.
%}

\iffalse %CCCCCCCCCCCCCCCCCCCCCCCCCCCCCCCCCCCCCCCCCCCCCCCCCCCCCCCCCCCCCCCCCC
\begin{definition}\label{def:confusable_set}
Suppose $\bx\in \Sigma_q^n$ is a string of length $n$. Let $B(\bx) \subseteq \Sigma_q^n$ denote the \emph{confusable set} of $\bx$ consisting of the largest subset of length-$n$ strings (excluding $\bx$) such that each $\by\in B(\bx)$ can generate the same output as $\bx$.
\end{definition}
\fi %CCCCCCCCCCCCCCCCCCCCCCCCCCCCCCCCCCCCCCCCCCCCCCCCCCCCCCCCCCCCCCCCCC

%\bcomment{
\begin{definition}\label{def:confusable_set}
A \emph{confusable set} $B(\bx)\subseteq S$  of $\bx\in S$ consists of all $\by\in S$, excluding $\bx$, such that $\bx$ and $\by$ can produce the same output when passed through the channel.
\end{definition}
%}

%Given $B(\bx)$ for $\bx\in \Sigma_{q}^n$, we introduce a \emph{labeling function}.

%\bcomment{
\begin{definition}\label{def:mapping}
Let $\cR(n)%\leq o((\log\log n)\cdot \log n )
$ be an integer function of  $n$. A \emph{labeling function for the confusable sets $B(\bx),\bx\in S,$} is a function
\[ f: \Sigma_q^n \to \Sigma_{2^{\cR(n)}} \]
such that, for any $\bx\in S$ and $\by \in B(\bx)$, $f(\bx)\neq f(\by)$.
\end{definition}

%\iffalse
%\begin{theorem}\label{Theo:syndrome}
%(c.f.\ \cite[Theorem~5]{sima2020syndrome}) Let $f:\Sigma_2^n\to \Sigma_{2^{\cR(n)}}$ be a labeling function with $\cR(n)\leq o((\log\log n)\cdot \log n )$ such that for any fixed $\bx\in \{0,1\}^{n}$ and any $\by \in B(\bx)$, we have $f(\bx)\neq f(\by)$. Then there exists an integer $a\leq 2^{\log \Vert B(\bx)\Vert+o(\log n)}$ such that for all $\by\in B(\bx)$, we have that $f(\bx) \not\equiv f(\by) \bmod a$.
%\end{theorem}
%\fi

\begin{theorem}\label{Theo:syndrome_qary}
(c.f.\ \cite[Theorem~5]{sima2020syndrome}) Let $f: \Sigma_q^n \to \Sigma_{2^{\cR(n)}}$, where $\cR(n)=o(\log\log n\cdot \log n )$, be a labeling function for the confusable sets $B(\bx),\bx\in S$. Then there exists an integer $a\leq 2^{\log \Vert B(\bx)\Vert+o(\log n)}$ such that for all $\by\in B(\bx)$, we have  $f(\bx) \not\equiv f(\by) \bmod a$.
\end{theorem}

%Due to the advantage of low redundancy, we construct error-correcting codes for $\leq \!\!3$-TDs and at most $p$ substitutions by applying the syndrome compression technique \cite{sima2020syndrome}.
%Given a string $\bx \in \Sigma_{q}^{n}$, we assume that $B(\bx)$, $f(\bx)$, and $a$ for $\leq \!\!3$-TDs and at most $p$ substitutions are available. Then the general form of the codeword can be represented as $(\bx, \br)$, where the appended string $\br$ consists of two parts of information: i) The value of the parameter $a$ and ii) $f(\bx) \mod a$. It requires a brute force search for the parameter $a$. A more detailed description of the codes for duplications and substitutions is presented in Section~\ref{subsec:framework}.

The above definitions and  theorem are used in our code construction based on syndrome compression, presented in Section~\ref{sec:ECC}. The construction and analysis rely on the confusable sets for the channel, discussed in the next section.

%>>>>>>>>>>>>>>>>>>>>>>>>>>>>>>>>>>>>>>>>>>>>>>>>>>>>>>>>>>>>>>>>>>>>>>>>>>>>>>>>

%%%%%%%%%%%%%%%%%%%%%%%%%%%%%%%%%%%%%%%%%%%%%%%%%%%%%%%%%%%%%%%%%%%%%%%%%%%%%%%%%%%%%%%%%%%%%%%%%%%%%%%%%%%%%%%
%%%%%%%%%%%%%%%%%%%%%%%%%%%%%%%%%%%%%%%%%%%%%%%%%%%%%%%%%%%%%%%%%%%%%%%%%%%%%%%%%%%%%%%%%%%%%%%%%%%%%%%%%%%%%%%

\section{Confusable sets for channels with short duplication and substitution errors}\label{Sec:channel_model}

In this section, %we study the size of the set $B_{\ir}^{\leq p}(\bx)$,
we study the size of confusable sets of input strings passing through channels with an arbitrary number of duplications and at most $p$ substitutions. This quantity will be used to derive a Gilbert-Varshamov bound and, in the next section, to construct our error-correcting codes.

Since the duplication root is unique, and duplications and deduplications do not alter the duplication root of the input,  $\ir_q(n)$ is a code capable of correcting duplications. The decoding process simply removes all tandem repeats.  %and this approach has been shown to produce  codes with asymptotically optimal rate~\cite{jain2017duplication,kovavcevic2019codes,chee2020efficient}.
In other words, if we append a root block, which deduplicates all repeats and produces the root of its input, to the channel with duplication errors, any irreducible sequence passes through this concatenated channel with no errors. This approach produces  codes with the same asymptotic rate as that of~\cite{jain2017duplication}, achieving the highest known asymptotic rate.

\tikzstyle{block} = [draw, fill=blue!10, rectangle, minimum height=2.5em, minimum width=4em]
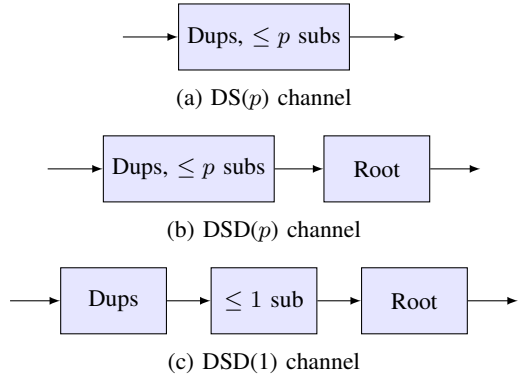
\begin{figure}
     \centering
     \begin{subfigure}[b]{\columnwidth}
         \centering
\begin{tikzpicture}[node distance=2cm]
    \node [] (input) {};
    \node [block, right of=input] (dsp) {Dups, $\le p$ subs};
    \node [right of=dsp] (output) {};
    \draw [draw,->] (input) -- (dsp);
    \draw [draw,->] (dsp) -- (output);
\end{tikzpicture}
    \caption{DS($p$) channel}
    \label{fig:DS}
     \end{subfigure}
\begin{subfigure}[b]{\columnwidth}
         \centering
     \vspace{.3cm}
\begin{tikzpicture}[node distance=2cm]
    \node [] (input) {};
    \node [block, right of=input] (dsp) {Dups, $\le p$ subs};
    \node [block, right of=dsp, xshift=.5cm] (dd) {Root};
    \node [right of=dd,xshift=-.5cm] (output) {};
    \draw [draw,->] (input) -- (dsp);
    \draw [draw,->] (dsp) -- (dd);
    \draw [draw,->] (dd) -- (output);
\end{tikzpicture}
    \caption{DSD($p$) channel}
    \label{fig:DSDp}
     \end{subfigure}
\begin{subfigure}[b]{\columnwidth}
         \centering
         \vspace{.3cm}
\begin{tikzpicture}[node distance=2cm]
    \node [] (input) {};
    \node [block, right of=input,xshift=-.5cm] (d) {Dups};
    \node [block, right of=d] (sub) {$\le 1$ sub};
    \node [block, right of=sub] (dd) {Root};
    \node [right of=dd,xshift=-.5cm] (output) {};
    \draw [draw,->] (input) -- (d);
    \draw [draw,->] (d) -- (sub);
    \draw [draw,->] (sub) -- (dd);
    \draw [draw,->] (dd) -- (output) {};
\end{tikzpicture}
    \caption{DSD(1) channel}
    \label{fig:DSD1}
     \end{subfigure}
     \caption{Any error-correcting code for channel (b) is also an error-correcting code for channel (a). The confusable set for a channel obtained by concatenating  $p$ copies of channel (c) contains the confusable set for channel (b). %\rcomment{Yuanyuan: why not replace "Root" by "deduplications"?}
     }
     \label{fig:chans}
\end{figure}

Similar to~\cite{Tang2021ECC_Edit}, we extend this strategy to design codes for the channel with duplication and at most $p$ substitution errors, denoted the DS($p$) channel and shown in Figure~\ref{fig:DS}. Note that the duplications and substitutions can occur in any order. We take the code to be a subset of irreducible strings and find the code for a new channel obtained by concatenating a root block to the channel with duplication and substitution errors, denoted as the DSD($p$) channel and shown in Figure~\ref{fig:DSDp}. Clearly, any error-correcting code for DSD($p$) is also an error-correcting code for the DS($p$) channel.%(recall that duplications and substitution errors can occur in any order).

We now define the confusable sets over $\ir_q(n)$ for the DSD($p$) channel and bound its size, which is needed to construct the code and determine its rate.
\begin{definition}\label{def:root_confusable_set}
For $\bx\in \ir_q(n)$, let
\begin{equation}
\begin{split}
      B_{\ir}^{\leq p}(\bx) =  \{ \by &\in \ir_q(n) : \by \neq \bx, \\
      &R(\ddes{\leq p}(\bx))\cap R(\ddes{\leq p}(\by))\neq \varnothing   \}
\end{split}
\end{equation}
denote the \emph{irreducible-confusable set} of $\bx$.
\end{definition}

%)))))))))))))))))))))))))))))))))))))))))))))))))))))))))))))))))))))))))))))))))

Note that the DSD(1) channel can be represented as shown in Figure~\ref{fig:DSD1}. This is because the sequence of errors consists of duplications, substitutions, more duplications, and finally all deduplications. Hence, duplications that occur after the substitutions are all deduplicated and we may equivalently assume they have not occurred. Next, observe that the confusable set for the concatenation of $p$ DSD(1) channels contains the confusable set for a DSD($p$) channel. We can thus focus on this concatenated channel. The advantage of considering DSD(1) is that it is reversible in the sense that if $\bv$ can be obtained from an input $\bu$, then $\bu$ can be obtained from the input $\bv$, and this simplifies our analysis.

%Since deduplications can be undone by duplications, %Since duplications and deduplications do not affect the roots of strings,
%instead of the DSD($p$) channel, we can consider a concatenation of $p$ DSD($1$) channels, without reducing the size of the confusable set. The input of each DSD($1$) channel suffers a number of duplications, at most one substitution, and then all possible deduplications.

\begin{figure}
    \centering
    \includegraphics[width=0.35\textwidth]{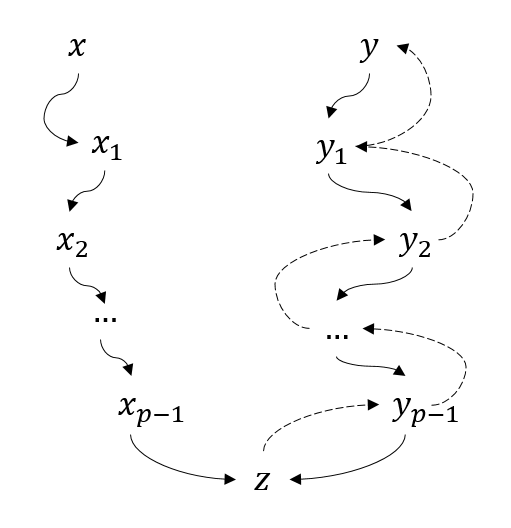}
    \caption{A sequence $\bz=\bx_p=\by_p$ that can be obtained from both $\bx$ and $\by$ through channels resulting from the concatenation of $p$ DSD(1) channels, each shown by a solid arrow. The dashed arrows represent the reverse relationships and each $\by_{i-1}$ can be obtained by passing $\by_i$ through a DSD(1) channel.% The equivalent channel by concatenating $p$ sub-channels of a channel with duplications, at most $p$ substitutions, and deduplications. Given $\bx\in \ir(n)$ and $\by \in  B_{\ir}^{\leq p}(\bx)$, let $\bz\in R(\ddes{\leq p}(\bx))\cap R(\ddes{\leq p}(\by))$. Furthermore, let $\bx_i \in R(\ddes{\leq i}(\bx))$ and $\by_i \in R(\ddes{\leq i}(\by))$ for $i\in[p]$ respectively. Note that $\bx_0=\bx$ and $\by_0=\by$, and $\bz=\bx_p=\by_p$.
    }
    \label{fig:dup_subs_ded_channel_and_equivalent_channel}
\end{figure}

Fig.~\ref{fig:dup_subs_ded_channel_and_equivalent_channel} shows a confusable string $\bz$ obtainable from irreducible sequences $\bx\in \ir_q(n)$ and $\by\in B_{\ir}^{\leq p}(\bx)$, after passing through $p$ DSD(1) channels, each  represented by a solid arrow. More precisely, $\bx_i \in R(\ddes{\leq 1}(\bx_{i-1}))$ and $\by_i \in R(\ddes{\leq 1}(\by_{i-1}))$, where $\bx=\bx_0,\by = \by_0, \bz = \bx_p=\by_p$. Furthermore, $\by_{i-1} \in R(\ddes{\leq 1}(\by_{i}))$. Hence, $\by$ can be generated from $\bx$ by concatenating the solid-line path from $\bx$ to $\bz$ and the dashed-line path from $\bz$ to $\by$, i.e., $\bx \to \bx_1 \to \dotsm\to \bz\to \by_{p-1} \to \dotsm\to \by$, where each $\to$ represents a DSD(1) channel. Considering the number of possibilities in each step gives the following lemma.

\begin{lemma}\label{lem:size_root_confusable_set_psubs_p_channels}
For $\bx \in \ir_q(n)$, \\
\resizebox{\linewidth}{!}{
  \begin{minipage}{\linewidth}
\begin{align*}
   &\Vert B_{\ir}^{\leq p}(\bx) \Vert \leq \max_{\bx_i, \by_j}  \prod_{i=0}^{p-1} \Vert  R(\ddes{\leq 1}(\bx_i)) \Vert \prod_{i=1}^{p} \Vert  R(\ddes{\leq 1}(\by_i)) \Vert\\
   %&\qquad\leq \prod_{i=0}^{p-1} \max_{\bx_i} \Vert  R(D^{*,\leq 1}(\bx_i)) \Vert \prod_{j=1}^{p} \max_{\by_j} \Vert  R(D^{*,\leq 1}(\by_j)) \Vert,
\end{align*}
\end{minipage}}
where the maximum for $\bx_i$ (resp.\ $\by_i$) is over sequences that can result from $\bx$ (resp.\ $\by$) passing through the concatenation of $i$ DSD(1) channels. %s are over $\bx_i \in R(\ddes{\leq i}(\bx))$, $\by_j \in R(\ddes{\leq j}(\by))$ with $\by\in B_{\ir}^{\leq p}(\bx)$, and $\bx_p=\by_p$.
\end{lemma}

%/////////////////////////////////////////////////////////////////////////

%According to Lemma~\ref{lem:size_root_confusable_set_psubs_p_channels}, finding  $\max_{\bx_i}  \Vert  R(\ddes{\leq 1}(\bx_i))\Vert$ and $\max_{\by_j} \Vert  R(\ddes{\leq 1}(\by_j))\Vert$ are the keys to deriving an upper bound on $\Vert B_{\ir}^{\leq p}(\bx) \Vert$. We can start by finding $\max_{\bx} \Vert  R(\ddes{\leq 1}(\bx))\Vert$ for some irreducible string $\bx$. Then we can extend it to $\max_{\bx_i}  \Vert  R(\ddes{\leq 1}(\bx_i))\Vert$ and $\max_{\by_j} \Vert  R(\ddes{\leq 1}(\by_j))\Vert$ for $\bx_i$ and $\by_j$ with bounded lengths, which will be discussed later.

It thus suffices to find  $\Vert R(\ddes{\leq 1}(\bx)) \Vert$. As
\begin{align*}%\label{eq:error_root_set}
    \Vert R(\ddes{\leq 1}(\bx)) \Vert &\leq \Vert R(\ddes{1}(\bx)) \Vert +\Vert R(\ddeo(\bx)) \Vert
    \\&= \Vert R(\ddes{1} (\bx)) \Vert+1,
\end{align*}
we find an upper bound on $\|R(\ddes{1} (\bx))\|$, in Lemma~\ref{lem:root_bound_by_abcde}, using the following lemma from \cite{Tang2021ECC_Edit}.
%////////////////////////////////////////////////////////////////////////////////////

\begin{lemma}\label{lem:decomposition}
\cite[Lemma~3]{Tang2021ECC_Edit}
Let $\bx$ be any string of length at least 5 and $\bx'\in \ddeo(\bx)$. %We can decompose $\bx$ and $\bx'$ as
For any decomposition of $\bx$ as
\[\bx=\ \br \ ab\ c \ de\ \bs,\]
for $a,b,c,d,e \in \Sigma_{q}$ and $\br,\bs \in \Sigma^{*}_{q}$, there is a decomposition of $\bx'$ as \[\bx'=\bu\  ab\ \bw \ de \ \bv\] such that $\bu, \bw,\bv \in \Sigma^{*}_{q}$,
 $\bu ab\in \ddeo(\br ab)$, $ab\bw de\in \ddeo(abcde)$, and $de\bv\in \ddeo(de\bs)$.
\end{lemma}

%Based on Lemma~\ref{lem:decomposition}, the upper bound of $\Vert R(\ddes{1} (\bx))\Vert$ for $\bx\in \ir(n)$ can be shown in the following lemma.

\begin{lemma}\label{lem:root_bound_by_abcde}
For an irreducible string $\bx \in \Sigma_q^n$, %with \rcomment{$n\geq 5$},
\begin{equation*}
    \Vert R(\ddes{1} (\bx)) \Vert\leq n \max_{\bt\in \Sigma^5_q} \Vert R(\ddes{1} (\bt))  \Vert.
\end{equation*}
\end{lemma}

%\iffalse
\begin{IEEEproof}
%Based on Lemma~\ref{lem:decomposition}, we explore the upper bound of $\Vert R(\ddes{1} (\bx))\Vert$ for $\bx\in \ir(n)$.
Given an irreducible string $\bx\in \ir_q(n)$, let $\bx'\in \ddeo(\bx)$ be obtained from $\bx$ through duplications and $\bx''$ obtained from $\bx' $  by a substitution. For a given $\bx$, $\Vert R(\ddes{1} (\bx)) \Vert$ equals the number of possibilities for $R(\bx'')$ as $\bx''$ varies. Note that duplications that occur after the substitution do not affect the root. So we have assumed that the substitution is the last error before the root is found.

Decompose $\bx$ as $\bx=\br abcde \bs$ with $\br,\bs\in\ir_q(*)$ and  $a,b,c,d,e\in\Sigma_{q}$, so that the substituted symbol in $\bx'$ is a copy of $c$. Note that if $|\bx|<5$ or if a copy of one of its first two symbols or its last two symbols is substituted, then we can no longer write $\bx$ as described. To avoid considering these cases separately, we may append two dummy symbols to the beginning of $\bx$ and two dummy symbols to the end of $\bx$, where the four dummy symbols are distinct and do not belong to $\Sigma_{q}$, and prove the result for this new string. Since these dummy symbols do not participate in any duplication, substitution, or deduplication events, the proof is also valid for the original $\bx$.

By Lemma~\ref{lem:decomposition}, we can write
\begin{equation}\label{eq:decomposition}
    \begin{split}
    \bx&=\ \br \ ab\ c \ de\ \bs\\
    \bx'&=\bu\  ab\ \bw \ de \ \bv% \in \ddeo(\bx)
    ,\\
    \bx'' &= \bu \ ab\ \bz \ de \ \bv, % \in D^{0,1}(\bx'),
    \end{split}
\end{equation}
where $\bu ab\in \ddeo(\br ab)$, $ab\bw de\in \ddeo(abc de)$,  $de\bv\in \ddeo(de\bs)$, and $\bz$ is obtained from $\bw$ by substituting a copy of $c$.
From~\eqref{eq:decomposition}, $R(\bx'')=R( \br R(ab \bz de ) \bs )$, where $R(ab \bz de )$ starts with $ab$ and ends with $de$ (which may fully or partially overlap).  %Note that both the substituted symbol and $c$ belong to $\Sigma_q$.

To determine $\|R(\ddes{1}(\bx))\|$, we count the number of possibilities for $R(\bx'')$ as $\bx''$ varies. Considering the decomposition of $\bx''$ into $\bu ab \bz de\bv$ given in~\eqref{eq:decomposition}, we note that if $R(ab\bz de)$ is given, then $R(\bx'')=  R( \br R(ab \bz de ) \bs )$ is uniquely determined. So to find an upper bound, it suffices to count the number of possibilities for $R(ab\bz de)$. We thus have
\[
\|R(\ddes{1}(\bx))\| \le \sum\|\{R(ab\bz de):ab \bz de \in\ddes{1}(abcde)\}\|,
\]
where the sum is over the choices of $c$ in $\bx$, or equivalently the decompositions of $\bx$ into $\br abcde\bs$, in~\eqref{eq:decomposition}. As there are $n$ choices for $c$, we have
\[
\|R(\ddes{1}(\bx))\| \le n\max_{\bt\in \Sigma^5_q} \Vert R(\ddes{1} (\bt))  \Vert.
\]
%For a fixed choice of $c$, the number of possibilities for some $ab \bz de \in\ddes{1}(abcde)$, the number of possibilities for $R(\bx'')$ is upper bounded by the number of possibilities for $ R(ab \bz de )$, which is in turn bounded by $\max_{\bt\in \Sigma^5_q} \Vert R(\ddes{1} (\bt))  \Vert$.}
%% for a given $\bu\in %%%\Sigma_q^5$. %In particular, applying deduplications among $ R( \br R(ab \bz de ) \bs ) $ for different $R(ab \bz de )$ may result in the same root, leading to a smaller number of distinct $R(\bx'')$ compared to the number of distinct $ R(ab \bz de )$. %Therefore, the number of distinct roots $ R(\bx'') =  R( \br R(ab \bz de ) \bs )$  by substituting a copy of $c$ is upper bounded by $\vert\vert  E_R^{*,1}(abcde) \vert\vert$, which represents the set of distinct roots $ R(ab \bz de )$ generated from $abcde \in \Sigma_{q}^5$.
%%%Based on \eqref{eq:decomposition}, let $\vert\vert \bar B_R^{*,1}(abcde) \vert\vert$ denote the set of roots generated from $abcde \in \Sigma_q^5$ by short duplications, a substitution of a copy of $c$, and deduplications.  Then we have $\max_{abcde}\vert\vert \bar B_R^{*,1}(abcde) \vert\vert \leq \max_{abcde}\vert\vert  B_R^{*,1}(abcde) \vert\vert$.
 %Furthermore, since there are $n$ choices of $c$, the size of $\Vert R(\ddes{1} (\bx))\Vert\leq n  \max_{\bu\in \Sigma^5_q} \Vert R(\ddes{1} (\bu))\Vert$, shown in the lemma.
 \end{IEEEproof}

% ////////////////////////////////////////////////////////////////////////////////////////////////

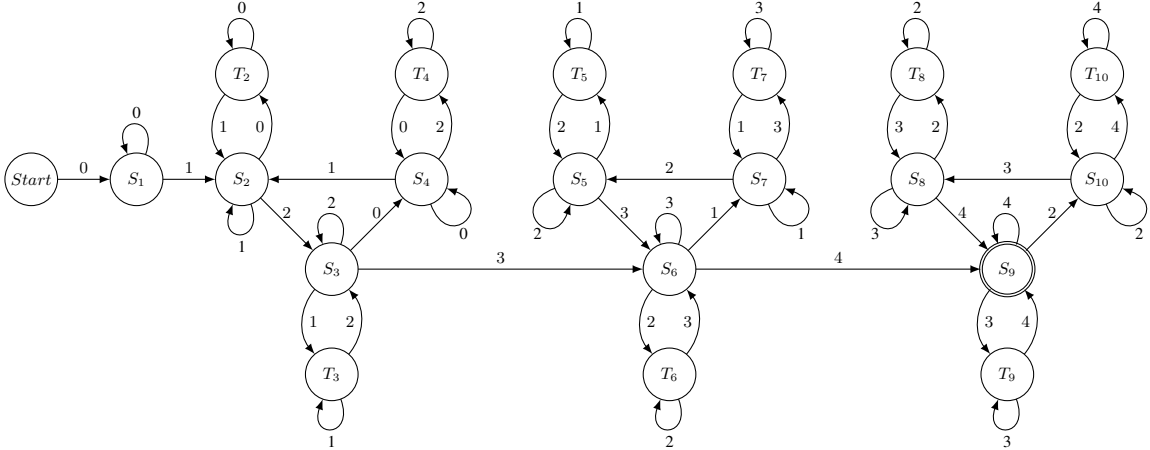
\begin{figure*}
    \begin{center}
        \begin{tikzpicture}[thin, scale=0.7, transform shape, vnd/.style={shape=circle,draw,inner sep=0,outer sep=0, minimum width=1cm, scale=1},
        invertibleedge/.style={->, line width=.3mm}]
            \node (Start) [vnd] {$Start$};
            \node (S1) [ right = of Start, vnd] {$S_1$};
            \node (S2) [ right = of S1, vnd] {$S_2$};
            \node (S3) [ below right = of S2, vnd] {$S_3$};
            \node (S4) [ above right = of S3, vnd] {$S_4$};
            \node (T2) [ above = of S2, vnd] {$T_2$};
            \node (T4) [ above = of S4, vnd] {$T_4$};
            \node (T3) [ below= of S3, vnd] {$T_3$};

            \node (S5) [ right = of S4, xshift=1cm, vnd] {$S_5$};
            \node (S6) [ below right = of S5, vnd] {$S_6$};
            \node (S7) [ above right = of S6, vnd] {$S_7$};
            \node (T5) [ above = of S5, vnd] {$T_5$};
            \node (T7) [ above = of S7, vnd] {$T_7$};
            \node (T6) [ below= of S6, vnd] {$T_6$};

            \node (S8) [ right = of S7, xshift=1cm,vnd] {$S_8$};
            \node[state,accepting] (S9) [ below right = of S8, vnd] {$S_9$};
            \node (S10) [ above right = of S9, vnd] {$S_{10}$};
            \node (T8) [ above = of S8, vnd] {$T_8$};
            \node (T10) [ above = of S10, vnd] {$T_{10}$};
            \node (T9) [ below= of S9, vnd] {$T_9$};

            \draw [->] (Start) -- (S1) node [midway, above, rotate=0] {$0$};
             \draw [->] (S1) -- (S2) node [midway, above, rotate=0] {$1$};
             \draw [->] (S2) -- (S3) node [midway, above, rotate=0] {$2$};
             \draw [->] (S3) -- (S4) node [midway, above, rotate=0] {$0$};
            \draw [->] (S4) -- (S2) node [midway, above, rotate=0] {$1$};

           \draw [->] (S1) edge [out=70,in=110,looseness=6] node[above] {0} (S1);
          \draw [->] (S2) edge [out=-70,in=-110,looseness=6] node[below] {1} (S2);
          \draw [->] (T2) edge [out=70,in=110,looseness=6] node[above] {0} (T2);
          \draw [->] (T4) edge [out=70,in=110,looseness=6] node[above] {2} (T4);
           \draw [->] (T3) edge [out=-70,in=-110,looseness=6] node[below] {1} (T3);

         \draw [->] (S3) edge [out=70,in=110,looseness=6] node[above] {2} (S3);
        \draw [->] (S4) edge [out=-70,in=-20,looseness=6] node[below] {0} (S4);

           % (S4) [out=150, in=20] to  (T4);
            \draw [->] (S4) to [out=50,in=-50]  node [midway, left, rotate=0]{$2$} (T4) ;
            \draw [->] (T4) to [out=230,in=130]  node [midway, right, rotate=0]{$0$} (S4);
            \draw [->] (S2) to [out=50,in=-50]  node [midway, left, rotate=0]{$0$} (T2) ;
            \draw [->] (T2) to [out=230,in=130]  node [midway, right, rotate=0]{$1$} (S2);
             \draw [->] (T3) to [out=50,in=-50]  node [midway, left, rotate=0]{$2$} (S3) ;
            \draw [->] (S3) to [out=230,in=130]  node [midway, right, rotate=0]{$1$} (T3);

             \draw [->] (S3) -- (S6) node [midway, above, rotate=0] {$3$};
             \draw [->] (S6) -- (S7) node [midway, above, rotate=0] {$1$};
             \draw [->] (S7) -- (S5) node [midway, above, rotate=0] {$2$};
            \draw [->] (S5) -- (S6) node [midway, above, rotate=0] {$3$};

            \draw [->] (T5) edge [out=70,in=110,looseness=6] node[above] {1} (T5);
          \draw [->] (T7) edge [out=70,in=110,looseness=6] node[above] {3} (T7);
           \draw [->] (T6) edge [out=-70,in=-110,looseness=6] node[below] {2} (T6);

        \draw [->] (S6) edge [out=70,in=110,looseness=6] node[above] {3} (S6);
        \draw [->] (S7) edge [out=-70,in=-20,looseness=6] node[below] {1} (S7);
        \draw [->] (S5) edge [out=200,in=250,looseness=6] node[below] {2} (S5);

           \draw [->] (S7) to [out=50,in=-50]  node [midway, left, rotate=0]{$3$} (T7) ;
            \draw [->] (T7) to [out=230,in=130]  node [midway, right, rotate=0]{$1$} (S7);
            \draw [->] (S5) to [out=50,in=-50]  node [midway, left, rotate=0]{$1$} (T5) ;
            \draw [->] (T5) to [out=230,in=130]  node [midway, right, rotate=0]{$2$} (S5);
             \draw [->] (T6) to [out=50,in=-50]  node [midway, left, rotate=0]{$3$} (S6) ;
            \draw [->] (S6) to [out=230,in=130]  node [midway, right, rotate=0]{$2$} (T6);

             \draw [->] (S6) -- (S9) node [midway, above, rotate=0] {$4$};
             \draw [->] (S9) -- (S10) node [midway, above, rotate=0] {$2$};
             \draw [->] (S10) -- (S8) node [midway, above, rotate=0] {$3$};
            \draw [->] (S8) -- (S9) node [midway, above, rotate=0] {$4$};

           \draw [->] (T8) edge [out=70,in=110,looseness=6] node[above] {2} (T8);
          \draw [->] (T10) edge [out=70,in=110,looseness=6] node[above] {4} (T10);
           \draw [->] (T9) edge [out=-70,in=-110,looseness=6] node[below] {3} (T9);

          \draw [->] (S9) edge [out=70,in=110,looseness=6] node[above] {4} (S9);
        \draw [->] (S10) edge [out=-70,in=-20,looseness=6] node[below] {2} (S10);
        \draw [->] (S8) edge [out=200,in=250,looseness=6] node[below] {3} (S8);

           \draw [->] (S10) to [out=50,in=-50]  node [midway, left, rotate=0]{$4$} (T10) ;
            \draw [->] (T10) to [out=230,in=130]  node [midway, right, rotate=0]{$2$} (S10);
            \draw [->] (S8) to [out=50,in=-50]  node [midway, left, rotate=0]{$2$} (T8) ;
            \draw [->] (T8) to [out=230,in=130]  node [midway, right, rotate=0]{$3$} (S8);
             \draw [->] (T9) to [out=50,in=-50]  node [midway, left, rotate=0]{$4$} (S9) ;
            \draw [->] (S9) to [out=230,in=130]  node [midway, right, rotate=0]{$3$} (T9);

            %\draw [invertibleedge] (S4) -- (T4) node [midway, above, rotate=0] {$2$};
            \end{tikzpicture}
    \caption{Finite automaton for the regular language $D^*(01234)$ based on~\cite{jain2017capacity}. } \label{fig:fsm5}
    \end{center}
\end{figure*}

% ////////////////////////////////////////////////////////////////////////////////////////////////

The next lemma provides a bound on  $\Vert R(\ddes{1} (\bt))\Vert$ for ${\bt\in\Sigma_q^5}$ by identifying the ``worst case''. The proof is given in Appendix~\ref{app:lem_dom}.

\begin{restatable}{lemma}{dominance}
\label{lem:dominance}
Given $q\geq 3$, we have
\begin{align*}
    \max_{\bt\in \Sigma^5_q} \Vert R(\ddes1(\bt))  \Vert\le \Vert R(\ddes{1} (01234))  \Vert,
\end{align*}
where $\ddes{1}(01234)\subseteq \Sigma_{q+4}^*$ (the substituted symbol can be replaced with another symbol from $\Sigma_{q+4}$).
\end{restatable}

As shown in~\cite{jain2017capacity}, $\ddeo (01234)$ is a regular language whose words can be described as paths from `Start' to $S_9$ in the finite automaton  given in Figure~\ref{fig:fsm5}. Then $\ddes1(01234)$ is equivalent to the set of paths from `Start' to $S_9$ but with the label on one edge substituted. We will use this observation to bound $\Vert R(\ddes{1} (01234))  \Vert$ in Lemma~\ref{lem:RD01234}. The next lemma establishes a symmetric property of the automaton that will be useful in Lemma~\ref{lem:RD01234}. Lemma~\ref{lem:symmetry} is proved by  showing that there is a bijective function $h:U\to V$ between $U$ and $V$ and between $R(U)$ and $R(V)$. Specifically, for $\bu=u_1\dotsm u_n$, we let $\bv = h(\bu) = \bar u_n \bar u_{n-1} \dotsc \bar u_{1}$, where for $a\in\{0,1,2,3,4\}$, $\bar a = 4-a$. A detailed proof is given in Appendix~\ref{app:symmetry}.

\begin{restatable}{lemma}{symmetry}
\label{lem:symmetry}
Let $U$ and $V$ be the sets of labels of all paths from Start to any state and from any state to $S_9$, respectively, in the finite automaton of Figure~\ref{fig:fsm5}. Then $\|U\|=\|V\|$ and $\|R(U)\|=\|R(V)\|$.
\end{restatable}

%The lemma is proved in Appendix~\ref{Appdix:lem:symmetry}.

\begin{lemma}\label{lem:RD01234}
For $\hat q\ge 5$ and $\ddes{1} (01234)\subseteq \Sigma^*_{\hat q}$, where the substitution replaces a symbol with any symbol from $\Sigma_{\hat q}$, we have
\[\Vert R(\ddes{1} (01234)) \Vert \leq 22^2 (\hat q-1).\]
\end{lemma}

\begin{IEEEproof}
Based on~\cite{jain2017capacity}, recall that $\ddeo (01234)$ is a regular language whose words can be described as paths from `Start' to $S_9$ in the finite automaton given in Figure~\ref{fig:fsm5}, where the word associated with each path is the sequence of the edge labels. Let $\bx'\in \ddeo(01234)$ and let $\bx''$ be generated from $\bx'$ by a substitution. Assume $\bx'=\bu w \bv$ and $\bx''=\bu \hat w \bv$, where $\bu ,\bv\in\Sigma_5^*,w\in\Sigma_5$ and %\rcomment{$\hat w \neq w \in \Sigma_{\hat q}$}
$\hat w\in \Sigma_{\hat q}\setminus\{w\}$. So there are  $\hat q-1$ choices for  $\hat w$. The string $\bu$ represents a path from `Start' to some state $s_u$ and the string $\bv$ represents a path from some state $s_v$ to $S_9$ in the automaton, where there is an edge with label $w$ from $s_u$ to $s_v$.

As $\bx''=\bu \hat w \bv$,  we have $R(\bx'') = R(R(\bu)\hat w R(\bv))$, where $R(\bu)$ is an irreducible string represented by a path from ``Start" to state $s_u$, and $R(\bv)$ is an irreducible string represented by a path from $s_v$ to $S_9$. Define $U$ and $V$ as in Lemma~\ref{lem:symmetry}. We thus have  %Lemma~\ref{lem:symmetry}, we have $R(\bu)\in R(U)$ and $R(\bu)\in R(V)$. Since $R(\bx'')$ is a singleton, we have
$\Vert R(\ddes{1} (\bx)) \Vert \leq \vert\vert R(U) \vert\vert \times (\hat q-1) \times \vert\vert R(V) \vert\vert = \vert\vert R(U) \vert\vert^2 \times (\hat q-1)$. By inspection, we can show that
\begin{align*}
      R(U)&=\{
     \Lambda,0,01,01201,012,0120,010,012010,\\ &0121,01202, 0123,01232,01231,012313,012312,\\
     &0123121, 01234,012343,012342,0123424,\\
     &0123423,01234232\},
\end{align*}
and hence $\vert\vert R(U) \vert\vert=22$, completing the proof.
\iffalse% \\\\\\\\\\\\\\\\\\\\\\\\\\\\\\\\\\\\\\\\\\\\\\\\\\\\\\\\\\\\\\\\\\\\\\\\\\\\\\\\\\\\\\\\\\\\\\\\
Based on Lemma~\ref{lem:symmetry}, $R(U)$ and $R(V)$ consist of all irreducible strings produced as labels of paths from Start to any state and from any state to $S_9$, respectively. The irreducible strings represented by paths from Start to each state and from each state to $S_9$ are listed in the second and the third columns of Table~\ref{Table:Irreducible_Paths_01234}, respectively. Then the two sets $R(U)$ and $R(V)$ are
 \begin{align*}
      R(U)&=\{
     \Lambda,0,01,01201,012,0120,010,012010,\\ &0121,01202, 0123,01232,01231,012313,012312,\\
     &0123121, 01234,012343,012342,0123424,\\
     &0123423,01234232\}, \\
     R(V)&=\{\Lambda,4,34,234,24234,4234,34234,434,\\
     & 434234,3234,21234, 1234,231234,31234,321234,\\
     &131234,101234, 01234,1201234,201234,\\
     &21201234,0201234\}.
 \end{align*}
 Then we have $\Vert R(U)\Vert =\vert R(V)\Vert=22$.  It is also easy to verify that $h(R(\bu))\in R(V)$ for each $R(\bu)\in R(U)$.
%Recall that $ \|R(U)\|=\| R(V)\|=22$ with $g(R(\bu))\in R(V)$ for every $R(\bu)\in R(U)$ in Lemma~\ref{lem:symmetry}.
Therefore, given $q\geq 3$ and $\hat q= q+4$, we have $\Vert R(\ddes{1} (01234)) \Vert \leq \|R(U)\| \times (\hat q-1) \times \|R(V)\| \leq 22^2 (q+3)$.

As a result, given $q\geq 3$, we have $\max_{\bx\in \Sigma_q^5}\Vert R(\ddes{1} (\bx)) \Vert \leq \Vert R(\ddes{1} (01234)) \Vert \leq 22^2 (q+3) \leq 22^2 \times 2q=968q$.
\fi% ////////////////////////////////////////////////////////////////////////////////////////////////
\end{IEEEproof}

% ////////////////////////////////////////////////////////////////////////////////////////////////

%\bcomment{
\begin{theorem}\label{Lem:max_root_set_upper_1subs}
For an irreducible string $\bx \in \Sigma^n_q$, with $q\geq 3$,
\begin{equation*}
    \Vert R(\ddes{\le1} (\bx)) \Vert\leq  968 nq+1.
\end{equation*}
%Then the size of the irreducible-confusable set of $\bx$ satisfies
%\[\vert\vert B_{\ir}^{*,1}(\bx) \vert\vert \leq \vert\vert E_R^{*,1}(\bx) \vert\vert^2\leq 968^2 q^2 n^2.\]
%Furthermore, $\Vert R(\ddes{\leq 1}(\bx))  \Vert \leq  \Vert R(\ddes{1} (\bx)) \Vert +1\leq  968 n q+1$.
\end{theorem}
%}
\begin{IEEEproof}
From Lemmas~\ref{lem:root_bound_by_abcde},~\ref{lem:dominance}, and~\ref{lem:RD01234}, it follows that $\Vert R(\ddes{1} (\bx)) \Vert\leq 22^2n(\hat q-1)\le  2q \cdot  22^2n=968nq$ with $\hat q = q+4$. Furthermore, $\Vert R(\ddes{\leq 1}(\bx))  \Vert \leq  \Vert R(\ddes{1} (\bx)) \Vert +1$.
\end{IEEEproof}

We can now use Theorem~\ref{Lem:max_root_set_upper_1subs} along with Lemma~\ref{lem:size_root_confusable_set_psubs_p_channels}, to find a bound on $\Vert B_{\ir}^{\leq p}(\bx) \Vert$. To do so, we need to bound the size of $\bx_i$ and $\by_i$ shown in Figure~\ref{fig:dup_subs_ded_channel_and_equivalent_channel}, for which the following theorem is of use. The theorem is a direct extension of \cite[Theorem~5]{Tang2021ECC_Edit} and thus requires no proof. An example demonstrating the theorem is given in Appendix~\ref{app:example}.

\begin{restatable}[c.f.{\cite[Theorem~5]{Tang2021ECC_Edit}}]{theorem}{rootpL}\label{Theo:root_L_17}%
Given strings $\bx \in \Sigma_q^n$ and $\bv\in \ddes{\le p}(\bx)$, $R(\bv)$ can be obtained from $R(\bx)$ by at most $p$ $\cL$-substring edits, where $\cL=17$.
\end{restatable}

It follows from the theorem that for $1\le i\le p$,%Based on Theorem~\ref{Theo:root_L_17}, the effect of  duplications and one substitution on the duplication roots can be viewed as a $\cL$-substring edit. By combing Theorem~\ref{Lem:max_root_set_upper_1subs} and~\ref{Theo:root_L_17}, since $\bx_i$ (resp.\ $\by_i$) is over sequences that can result from $\bx$ (resp.\ $\by$) passing through the concatenation of $i$ DSD(1) channels, %given $\bx_i\in R(D^{*,\leq i}(\bx))$ and $\by_j\in R(D^{*,\leq j}(\by))$ with $i,j\leq p$,
%we have
\begin{equation}\label{eq:length_roots}
    |\bx_i| \leq n+p\cL,\ |\by_i| \leq n+p\cL.
\end{equation}
%By combining Lemma~\ref{lem:size_root_confusable_set_psubs_p_channels}, Lemma~\ref{Lem:max_root_set_upper_1subs}, and eq~\eqref{eq:length_roots}, the size of $\Vert B_{\ir}^{\leq p}(\bx) \Vert$ can be shown in the following theorem.
The next theorem then follows from Lemmas~\ref{lem:size_root_confusable_set_psubs_p_channels} and~\ref{Lem:max_root_set_upper_1subs}.

\begin{theorem}\label{Theo:max_root_set_p_subs}
Let $\bx \in \ir_q(n)\subseteq \Sigma^n_q$ be an irreducible string of length $n$ with $q\geq 3$. The irreducible-confusable set $B_{\ir}^{\leq p}(\bx)$ of $\bx$ satisfies
\begin{align*}
    \Vert B_{\ir}^{\leq p}(\bx) \Vert  \leq  (968q(n+p\cL)+1)^{2p}.
\end{align*}
\end{theorem}

%\textcolor{blue}{
The size of the confusable sets will be used for our code construction. It also allows us to derive a Gilbert-Varshamov (GV) bound, as follows.

\begin{theorem}\label{thm:GV-bound}
There exists a code of length $n$ capable of correcting any number of duplications and at most $p$ substitutions with size at least
\[
\frac{\|\ir_q(n)\|}{(968q(n+p\cL)+1)^{2p}}\cdot
\]\end{theorem}
We will show in Lemma~\ref{lem:dup_codes_len_n} that the size of the code with the highest asymptotic rate for correcting duplications only is essentially $\|\ir_q(n)\|$. Assuming that $p$ and $q$ are constants, this GV bound shows that a code exists for additionally correcting up to $p$ substitution errors with extra redundancy of approximately $2p\log_q n$ symbols. The two constructions presented in the next section have extra redundancies of $4p\log_qn$ and $8p\log_qn$, which are only small constant factors away from this existential bound.
%}

%%%%%%%%%%%%%%%%%%%%%%%%%%%%%%%%%%%%%%%%%%%%%%%%%%%%%%%%%%%%%
%%%%%%%%%%%%%%%%%%%%%%%%%%%%%%%%%%%%%%%%%%%%%%%%%%%%%%%%%%%%%%%%%
%%%%%%%%%%            Section IV
%%%%%%%%%%%%%%%%%%%%%%%%%%%%%%%%%%%%%%%%%%%%%%%%%%%%%%%%%%%%%%%%%
%%%%%%%%%%%%%%%%%%%%%%%%%%%%%%%%%%%%%%%%%%%%%%%%

%\clearpage
\section{Low-redundancy error-correcting codes}\label{sec:ECC}

As stated in Section~\ref{Sec:channel_model}, our code for correcting duplications and substitutions is a subset of irreducible strings of a given length.  In this section, we construct this subset by applying the syndrome compression technique~\cite{sima2020syndrome}, where we will make use of the size of the irreducible-confusable set $\Vert B_{\ir}^{\leq p}(\bx) \Vert$ derived in Section~\ref{Sec:channel_model}. In this section, unless otherwise stated, we assume that both $q\geq 4$ and $p$ are constant.

We begin by presenting the code constructions for correcting duplications and \textit{substitutions} in Subsection~\ref{subsec:framework}, assuming the existence of appropriate labeling functions used to produce the syndrome information and an auxiliary error-correcting code used to protect it. The labeling functions will be discussed in Subsection~\ref{subsec:labeling_function}, while the auxiliary ECC is presented in Section~\ref{Sec:Aux_ECC}. In Subsection~\ref{ssc:generalize}, we show that the proposed codes can in fact correct duplications and \textit{edits}. The redundancy of the codes and the computational complexities of their encoding and decoding are discussed in Subsections~\ref{subsec:codes_and_redundancy} and~\ref{secIV:time_complexity}, respectively. %Then, we derive expressions for the redundancy of the resulting codes, followed by an analysis of the encoding and decoding time complexities.
%We begin by first describing the general structure for each of our codewords. Based on this general description, in Section~\ref{subsec:decoding} a decoding algorithm is presented. Afterwards, we identify a labeling function that is compatible with the decoding algorithm discussed in Section~\ref{subsec:decoding}. Then, we derive expressions for the redundancy of the resulting codes, followed by the encoding process. Finally, the decoding/decoding time complexities are analyzed.

%%%%%%%%%%%%%%%%%%%%%%%%%%%%%%%%%%%%%%%%%%%

\subsection{Code constructions}\label{subsec:framework}
We first present a construction that assumes an error-free side channel is available, where the length of the sequence passing through the side channel is logarithmic in the length of the sequence passing through the main channel. In DNA storage applications, an error-free side channel may be available, for example, through data storage in silicon-based devices. We then present a second construction that does not make such an assumption, using a single noisy channel. In addition to potential practical uses, the first construction also helps make the second construction more clear by motivating some of its components.
%In this subsection, we will present the code construction.
%We will start from a preliminary code, given in~\eqref{eq:code1}, and address its shortcomings, building up to the final code given in Construction~\ref{const:code}. %first present the general structure for the codewords in our duplication and substitution error-correcting code. Note that the error-correcting code is a subset of irreducible strings of a given length.

\subsubsection{Channels with error-free side channels}
%Note that, for simplicity, $a$ and $f(\bx) \bmod a$ are considered as both integers and binary strings.
In the construction below, $\bx$ is transmitted through the noisy channel, while $\br$, which  encodes the information $(a,f(\bx) \bmod a)$ is transmitted through an error-free channel.

\begin{construction}\label{const:side-channel}
    Let $n,p,q$ be positive integers. Furthermore, let $f$ be a (labeling) function and, for each $\bx\in \ir_q(n)$, $a_\bx$ be a positive integer, such that
%$f(\bx)\neq f(\by)$ for $\by \in B_{\ir}^{\leq p}(\bx)$. Based on Theorem~\ref{Theo:syndrome}, we can generate an $a$, which is at most $2^{\log \Vert B_{\ir}^{\leq p}(\bx)\Vert+o(\log n)}$, such that
for any $\by \in B_{\ir}^{\leq p}(\bx)$, $f(\bx)\not \equiv f(\by) \bmod a_\bx$. Define%
        \begin{equation*}\label{eq:code1}
    \codeA_n=\{ (\bx,\br) : \bx \in \ir_q(n), \br = (a_\bx,f(\bx)\bmod a_\bx) \},
\end{equation*}
where $\br$ is assumed to be the $q$-ary representation of $(a_\bx,f(\bx)\bmod a_\bx)$.
\end{construction}
We consider the length of this code to be $N=n+|\br|$. As will be observed in~\eqref{eq:r-CA}, $|\br|=O(\log_q n)$ and so the sequence carried by the side channel is logarithmic in length. Recall that the existence of the labeling functions is discussed in  Subsection~\ref{subsec:labeling_function}.

\begin{theorem}
The code in Construction~\ref{const:side-channel}, assuming the labeling function $f$ and $a_\bx$ (for each $\bx\in\ir_q(n)$) exist, can correct any number of duplications and at most $p$ substitutions applied to $\bx$, provided that $\br$ is transmitted through an error-free channel.
\end{theorem}
\begin{IEEEproof}
Let the retrieved word from storing $\bx$ be $\bv\in R(\ddes{\le p}(\bx))$. Note that $a_\bx$ and $f(\bx)\bmod a_\bx$ can be recovered error-free from $\br$. By definition, for all $\by\neq \bx$ that could produce the same $\bv$, we have $\by \in B_{\ir}^{\leq p}(\bx)$. But then, $f(\by)\not\equiv f(\bx)\bmod a_\bx$, and so we can determine $\bx$ by exhaustive search.
\end{IEEEproof}
%Given $\bx\in \ir(n)$, let $\bz\in \ddes{p}(\bx)$ be an output of the DSD($p$) channel. According to Theorem~\ref{Theo:root_L_17}, $\bz$ can be considered to be generated from $\bx$ by $2p\cL$ insertions and deletions. Then a labeling function $f$ for deletion and insertion channels also works for the DSD($p$) channel. More discussion of the labeling function is presented in Subsection~\ref{subsec:labeling_function}.

\subsubsection{Channels with no side channels}
To better illustrate the construction with no side channels, let us first observe what the issues are with simply concatenating $\bx$ and $\br$ and forming codewords of the form~$\bx\br$.
\begin{itemize}
    \item The code in Construction~\ref{const:side-channel} relies on a sequence $\bv\in R(\ddes{\le p}(\bx))$ but if $\bx\br$ is stored, the output of the channel is a sequence $\bw\in R(\ddes{\le p}(\bx\br))$. As the boundary between $\bx$ and $\br$ becomes unclear after duplication and substitution errors, it is difficult to find $\bv\in R(\ddes{\leq p}(\bx))$ from $\bw\in R(\ddes{\leq p}(\bx\br))$. To address this, instead of finding $\bv$, we find a sufficiently long prefix, as discussed in Lemma~\ref{lem:prefix}. This will also require us to modify the labeling function.
    \item The decoding process requires the information encoded in $\br$, which is now subject to errors. We will address this by using a high-redundancy code that can protect this information, introduced in Lemma~\ref{lem:last_paper} and discussed in detail in Subsection~\ref{subsec:proof_lemma}.
    \item The codewords need to be irreducible. This is discussed in Lemma~\ref{lem:buffer}.
\end{itemize}

%is a code capable of correcting duplications and at most $p$ substitutions provided that, given the output $\bw \in R(\ddes{\leq p}(\bx\br))$, the following conditions are satisfied: 1) we can recover $(a,f(\bx) \bmod a)$ and 2) we can find some $\bv \in R(\ddes{\leq p}(\bx))$. To see this, observe that if $\by\neq \bx$ can also produce $\bv$, then $\by\in B^{\le p}_{\ir}(\bx)$, and hence it can be eliminated as an input candidate since $f(\by)\not\equiv f(\bx) \bmod a$. %Note that each $\bx\br$ is assumed to be irreducible since our error-correcting code is a subset of irreducible strings of a given length. The operation to ensure that the concatenation of $\bx$ and $\br$ is irreducible will be presented in Lemma~\ref{lem:buffer}.

For integers $p,j$, denote by $\ddesded{\leq p}{\leq j}(\bx)$ the set of strings that can be obtained by deleting a suffix of length at most $j$ from some $\bv \in R(\ddes{\leq p}(\bx))$. Note that $\ddesded{\leq p}{\leq j}(\bx)\subseteq\ir_q(*).$ %\rcomment{Not sure why it is important to say that the set consists of irreducible sequences. But if we want to say it, it shouldn't be part of the definition. It is a property.}

\begin{restatable}{lemma}{prefix}\label{lem:prefix}
     Let $\bx$ be an irreducible string of length $n$ and $\br$ any string such that $\bx\br$ is irreducible. Let $\bw\in R(\ddes{\leq p}(\bx\br))$ and $\bs$ be the prefix of $\bw$ of length $n-p\cL$. Then  %$\bs$ can be obtained from some $\bv\in R(\ddes{\leq p}(\bx))$ by deleting a suffix of length at most $2p\cL$. That is,
     $\bs\in \ddesded{\leq p}{\leq 2p\cL}(\bx)$.
\end{restatable}
The lemma is proved in Appendix~\ref{app:prefix}.

By choosing the first $n-p\cL$ elements of $\bw \in R(\ddes{\leq p}(\bx\br))$, we find $\bs\in \ddesded{\leq p}{\leq 2p\cL}(\bx)$, which is a function of only $\bx$ rather than $\bx\br$. But in doing so, we have introduced an additional error, namely deleting a suffix of length at most $2p\cL$. As a result, we need to  replace the labeling function $f$ with a stronger labeling function $f'$ that, in addition to handling both substitutions and duplications, can handle deleting a suffix of $\bx$. More precisely, $f'$ is a labeling function for the confusable set
\begin{equation}\label{eq:new-confusable}
\begin{split}
    &B^{\leq p,\leq 2 p\cL}_{\ir}(\bx)=\{\by \in \ir_q(n): \\
    &\quad\by\neq \bx, \ddesded{\leq p}{\leq 2p\cL}(\bx) \cap  \ddesded{\leq p}{\leq 2p\cL}(\by)\neq \varnothing\}.
    \end{split}
\end{equation}
The details of determining $f'$ will be discussed in Section~\ref{subsec:labeling_function}. Assuming the existence of the labeling function, $\br$ encodes $(a'_\bx,f'(\bx)\bmod a'_\bx)$, where for $\bx\in\ir_q(\bx)$, $a'_\bx$ is chosen such that \[f'(\bx)\not\equiv f'(\by)\bmod a'_\bx, \forall \by\in B^{\leq p,\leq 2 p\cL}_{\ir}(\bx).\]

To address the second difficulty raised above, i.e., protecting the information encoded in $\br$, we use an auxiliary high-redundancy code given in Section~\ref{Sec:Aux_ECC}. The following lemma, which is proved in Subsection~\ref{subsec:proof_lemma}, provides an encoder for this code.

\begin{restatable}{lemma}{encoder}
\label{lem:last_paper}
Let $\bsigma = 01020$. There exists an encoder $\cE_1:\Sigma_2^{L}\to \ir_q(L')$ such that i) $\bsigma\cE_1(\bu)\in\ir_q(*)$ and ii) for any string $\bx\in \ir_q(*)$ with $\bx\bsigma\cE_1(\bu)\in \ir_q(*)$, we can recover $\bu$  from any $\bw\in R(\ddes{\leq p}(\bx\bsigma\cE_1(\bu)))$. %In other words, $\bsigma \cE_1(\bu)$ can correct  prepending any irreducible string followed by duplications and at most $p$ substitutions.
Asymptotically, $L'\le \frac L{\log(q-2)}(1+o(1))$.
\end{restatable}

%\rcomment{Yuanyuan: since $\log $ represents $\log_q$, we need use $\log_2$ here!}

We use $\cE_1(a'_\bx,f'(\bx)\bmod a'_\bx)$ to denote $\cE_1(\bu)$, where $\bu$ is a binary sequence representing the pair $(a'_\bx,f'(\bx)\bmod a'_\bx)$. For $\bx\in \ir_q(n)$, by letting $\br=\cE_1(a'_\bx,f'(\bx)\bmod a'_\bx)$, we can construct codewords of the form $\bx\bsigma\br$. But such codewords would not necessarily be irreducible. Irreducibility can be ensured by adding a buffer $\bb_\bx$ between $\bx$ and $\bsigma\br$, as described by the next lemma, proved in Appendix~\ref{Appdix:lemma:length_buffer}.

\begin{restatable}{lemma}{buffer}\label{lem:buffer}
    For $q\ge 3$ and any irreducible string $\bx$ over $\Sigma_q$, there is a string $\bb_\bx$ of length $c_q$ such that $\bx\bb_\bx\bsigma$ is irreducible. Furthermore, $c_3= 13$, $c_4= 7$, $c_5=6$, and $c_q=5$ for $q\geq 6$.
\end{restatable}
\iffalse
\begin{lemma}\label{lem:buffer}
    For $q\ge 3$ and any irreducible string $\bx$ over $\Sigma_q$, there is a string $\bb_\bx$ of length $c_q$ such that $\bx\bb_\bx\bsigma$ is irreducible. Furthermore, $c_3= 13$, $c_4= 7$, and $c_q= 6$ for $q\ge 5$.
\end{lemma}
\fi

%From the lemma, letting $\br = \bsigma\cE_1(a,f(\bx)\bmod a)$ will enable us to recover $(a,f(\bx)\bmod a)$ from any $\bw\in R(\ddes{\leq p}(\bx\br))$ for any $\bx\in\ir(*)$ provided that $\bx\br\in \ir(*)$. We will discuss ensuring $\bx\br$ is irreducible later.
%the information $a$ and $f(\bx) \bmod a$ can be successfully recovered from ${\bw}$. %, as stated in Lemma~\ref{lem:C_MDS}.
%However, the task of identifying some $\bv \in \ddes{\leq p}(\bx)$ from $\bw$ proves to be  more difficult since we cannot detect the tail of $\bv$ in $\bw$.

The lemma implies that $\bx\bb_\bx\bsigma\br$ is irreducible. This is because any substring of length at most 6 is either in $\bx\bb_\bx\bsigma$ or $\bsigma\br$ but cannot span both as $|\bsigma|=5$. But $\bx\bb_\bx\bsigma$ and $\bsigma\br$ are both irreducible, as shown in Lemma~\ref{lem:buffer} and Lemma~\ref{lem:last_paper}.i, respectively.% starts with $\bsigma$, which has length 5 and because since short tandem repeats have length at most 6. So any repeat must be contained in $\bx\bb_\bx$ or in $\br$, which is not possible.
%It can be shown, similar to~\cite[Theorem~15]{Tang2021ECC_Edit}, that for any $\bx$, there is a string $\bb_\bx$ of a predetermined length $c_q$, such that $\bx \bb_\bx\bsigma \in \ir(*)$. Since $\br$ starts with $\bsigma$, which has length 5, we have $\bx \bb_\bx\br\in\ir(*)$.

We are now ready to present the code construction.% and then a theorem that establishes its error-correcting capability. The proof of the theorem summarizes our preceding discussion.
\begin{construction}\label{const:code}
Let $f'$ be a labeling function for the confusable sets $B^{\leq p,\leq 2 p\cL}_{\ir}(\bx),\bx\in\ir_q(n)$. Furthermore, for each $\bx$, let $a'_\bx$ be an integer such that $f'(\bx)\not\equiv f'(\by)\bmod a'_\bx$ for $\by\in B^{\leq p,\leq 2 p\cL}_{\ir}(\bx)$. Let
\[
\codeB_n = \{\bx \bb_\bx\bsigma \br: \bx\in\ir_q(n), \br = \cE_1(a'_\bx, f'(\bx) \bmod a'_\bx )\}.\]
\end{construction}

Note that for simplicity, we index the code by the length of $\bx$ rather than the length of the codewords $\bx \bb_\bx \bsigma\br$, i.e., $n$ in $\codeB_n$ refers to the length of $\bx$. The length of $\br$ is discussed in Subsection~\ref{subsec:codes_and_redundancy} below.

\begin{theorem}\label{Theo:ecc_psubs}
The code in Construction~\ref{const:code} can correct any number of short duplications and at most $p$ substitutions.
\end{theorem}
%The proof of the theorem summarizes our preceding discussion.
\begin{IEEEproof}
Let the retrieved word be $\bw\in R(\ddes{\leq p}(\bx\bb_\bx\bsigma\br))$. From Lemma~\ref{lem:last_paper}, given $\bw$, we can find $a'_\bx$ and $f'(\bx) \bmod a'_\bx$. Let $\bs$ be the $(n-p\cL$)-prefix of $\bw$. By Lemma~\ref{lem:buffer}, $\bx\bb_\bx\bsigma\br$ is irreducible. Then, by Lemma~\ref{lem:prefix}, the $(n-p\cL$)-prefix of $\bw$, denoted $\bs$, satisfies $\bs\in \ddesded{\leq p}{\leq 2p\cL}(\bx)$. By definition, for all $\by\neq \bx$ that could produce the same $\bs$, we have $\by\in B_{\ir}^{\le p,\le 2p\cL}(\bx)$. But then, $f'(\by)\not\equiv f'(\bx)\bmod a'_\bx$, and so we can determine $\bx$ by exhaustive search.
\end{IEEEproof}

%\bcomment{
\subsection{Extension to edit errors} \label{ssc:generalize}
We now show that the codes in Constructions~\ref{const:side-channel} and~\ref{const:code} are able to correct an arbitrary number of duplications and at most $p$ \textit{edit} errors, where an edit error may be a deletion, an insertion, or a substitution.

Define the DED(1) and DED($p$) channels analogously to the DSD(1) and DSD($p$) channels by replacing substitutions with edit errors. Any error-correcting code for a concatenation of $p$ DED(1) channels is also an error-correcting code for DED($p$).

Additionally, any error-correcting code for a DSD(1) channel is also an error-correcting code for the DED(1) channel. This is because any input-output pair $(\bx,\by)$ for DED(1), shown in Figure~\ref{fig:DED1}, is also an input-output pair for the DSD(1) channel, shown in Figure~\ref{fig:DSD1-2}. This claim is proved in~\cite[Corollary~12]{Tang2021ECC_Edit}, where it was shown that a deletion can be represented as a substitution and a deduplication, e.g., $abc\to ac$ as $abc\to aac\to ac$, and an insertion as a duplication and a substitution, e.g., $abc\to abdc$ as $abc\to abbc\to abdc$.

Since $\codeA$ and $\codeB$ can correct errors arising from a concatenation of $p$ DSD(1) channels, they can also correct errors arising from a concatenation of $p$ DED(1) channels and thus a DED($p$) channel, leading to the following theorem.

\tikzstyle{block} = [draw, fill=blue!10, rectangle, minimum height=2.5em, minimum width=4em]
\begin{figure}
     \centering
\begin{subfigure}[b]{\columnwidth}
         \centering
         \vspace{.3cm}
\begin{tikzpicture}[node distance=2cm]
    \node [] (input) {};
    \node [block, right of=input,xshift=-.5cm] (d) {Dups};
    \node [block, right of=d] (sub) {$\le 1$ sub};
    \node [block, right of=sub] (dd) {Root};
    \node [right of=dd,xshift=-.5cm] (output) {};
    \draw [draw,->] (input) -- (d);
    \draw [draw,->] (d) -- (sub);
    \draw [draw,->] (sub) -- (dd);
    \draw [draw,->] (dd) -- (output) {};
\end{tikzpicture}
    \caption{DSD(1) channel}
    \label{fig:DSD1-2}
     \end{subfigure}
\begin{subfigure}[b]{\columnwidth}
         \centering
         \vspace{.3cm}
\begin{tikzpicture}[node distance=2cm]
    \node [] (input) {};
    \node [block, right of=input,xshift=-.5cm] (d) {Dups};
    \node [block, right of=d] (sub) {$\le 1$ edit};
    \node [block, right of=sub] (dd) {Root};
    \node [right of=dd,xshift=-.5cm] (output) {};
    \draw [draw,->] (input) -- (d);
    \draw [draw,->] (d) -- (sub);
    \draw [draw,->] (sub) -- (dd);
    \draw [draw,->] (dd) -- (output) {};
\end{tikzpicture}
    \caption{DED(1) channel}
    \label{fig:DED1}
     \end{subfigure}
     \caption{Any error-correcting code for channel (a) is also an error-correcting code for channel (b). %\rcomment{Yuanyuan: why not replace "Root" by "deduplications"?}
     }
     \label{fig:DSDDED}
\end{figure}
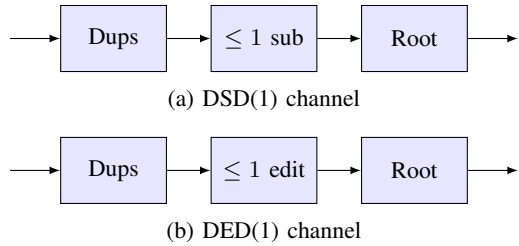
\begin{theorem}
The codes in Constructions~\ref{const:side-channel} and~\ref{const:code} can correct any number of duplications and at most $p$ edit errors.
\end{theorem}
%}

\iffalse% \\\\\\\\\\\\\\\\\\\\\\\\\\\\\\\\\\\\\\\\\\\\\\\\
\begin{IEEEproof}
Let DED($p$) channel represent a channel consisting of an arbitrary number of duplications, at most $p$ edit errors, and deduplicating all copies.  Then an code that corrects errors in the DED($p$) channel is also capable of correcting an arbitrary number of short duplications and at most $p$ edit errors.

Based on Section~\ref{Sec:channel_model}, the code in Construction~\ref{const:code} can correct errors in a channel that concatenates $p$ DSD($1$) channels. According to~\cite[Corollary~12]{Tang2021ECC_Edit}, each DED($1$) channel with duplications, at most one edit error, and deduplications can be viewed as a DSD($1$) channel. Therefore, the code in Construction~\ref{const:code} can also correct errors in a channel concatenating $p$ DED($1$) channels.  Since deduplications can be undone by duplications, the code in Construction~\ref{const:code} can correct errors in a DED($p$) channel. Hence, we extend Theorem~\ref{Theo:ecc_psubs} to correct edit errors.
\end{IEEEproof}
\fi% //////////////////////////////////////////////////////////////////////////////

%\rcomment{Update here!}

%%%%%%%%%%%%%%%%%%%%%%%%%%%%%%%%%%%
%%%%%%%%%%%%%%%%%%%%%%%%%%%%%%%%%%%%%
%%%%%%%%%%%%%%%%%%%%%%%%%%%%%%%%%%%%%%%%%

%%%%%%%%%%%%%%%%%%%%%%%%%%%%%%%%%%%%%%%%%%%%%%%%%%%%%%%%%

%%%%%%%%%%%%%%%%%%%%%%%%%%%%%%%%%%%%%%%%%%%%%%%%%%%%

\subsection{The labeling function} \label{subsec:labeling_function}

In this subsection, we first present the labeling function $f$ such that $f(\bx)\neq f(\by)$ for $\by \in B^{\leq p}_{\ir}(\bx)$, used in Construction~\ref{const:side-channel}. By Theorem~\ref{Theo:root_L_17}, $\bz\in R(\ddes{\leq p}(\bx)) \cap R(\ddes{\leq p}(\by))$ can be obtained from $\bx$ and from $\by$ by at most $2p\cL$ indels. Hence, it suffices to find $f$ such that $f(\bx)\neq f(\by)$ if there is a string $\bz$ that can be obtained from both $\bx$ and $\by$ through $2p\cL$ indels. Note that since we are utilizing syndrome compression, choosing a more ``powerful'' labeling function does not increase the redundancy, which is still primarily controlled by $\max_{\bx\in\ir_q(n)}\|B^{\leq p}_{\ir}(\bx)\|$. We use the next theorem for binary sequences to find $f$.
%Therefore a labeling function $f'$ working for the general insertion and deletion channel also works for the channel with duplications, substitutions, deduplications, and at most one suffix-burst deletion.

%To correct both deletions and insertions over $\Sigma_q$, we recall a labeling function $g$ that works for deletions, insertions, and substitutions over binary sequences~\cite{sima2020optimal_Systematic}. %\bcomment{Given $\bc\in \{0,1\}^n$, let $D_t(\bx)$ denote the set of strings generated from $\bc$ by at most $t$ insertions, deletions, and substitutions. Then we have the following lemma.}

\begin{theorem}\label{Theo:Hash_function_for_t_deletions}
\cite{sima2020optimal_Systematic} There exists a labeling function $g: \{0,1\}^{n}\to \Sigma_{2^{\cR(t,n)}}$ such that for any two distinct  strings $\bc_1$ and $\bc_2$ confusable under at most $t$ insertions, deletions, and substitutions, we have $g(\bc_1)\neq g(\bc_2)$%over $\bc$
, where $\cR(t,n)=[(t^2+1)(2t^2+1)+2t^2(t-1)]\log n+o(\log n)$.
\end{theorem}

Since $\bz\in  R(\ddes{\leq p}(\bx))$ can be obtained from $\bx$ via at most $2p\cL$ indels, $\cU_i(\bz)$ can be derived from $\cU_i(\bx)$ by at most $2p\cL$ indels, for $i\in [\lceil \log q \rceil]$.
Based on Theorem~\ref{Theo:Hash_function_for_t_deletions} and the work in~\cite{sima2020optimal_Systematic}, by letting $t=2p\cL$, we can obtain a labeling function $g$ for recovering $\cU_i(\bx)$ from $\cU_i(\bz)$ under at most $2p\cL$ indels.
Therefore, $f: \Sigma_{q}^n\to \Sigma_{2^{\lceil\log q\rceil \cR(t,n)}}$,
\begin{equation}\label{eq:labeling_function_q_arry}
    f(\bx)=\sum_{i=1}^{\lceil\log q\rceil} 2^{\cR(t,n)(i-1)}g(\cU_i(\bx)),
\end{equation}
where $t=2p\cL$, is a labeling function for the confusable sets $B^{\leq p}_{\ir}(\bx)$, $\bx\in \ir_q(n)$. For each $\bx$, a value $a_\bx$ needs to be also determined such that $f(\bx)\not\equiv f(\by)\bmod a_\bx$ for $\by\in B^{\leq p}_{\ir}(\bx)$. The existence of such $a_\bx$, satisfying $\log a_\bx\le {\log\|B^{\leq p}_{\ir}(\bx)\|+o(\log n)}$, is guaranteed by Theorem~\ref{Theo:syndrome_qary} provided that $p$ is a constant (ensuring that $p^4=o(\log \log n)$). The labeling function $f$ and integers $a_\bx$ are used in Construction~\ref{const:side-channel}.

In a similar manner, we can construct $f'$ as a labeling function for $B^{\leq p,\leq 2 p\cL}_{\ir}(\bx),\bx\in\ir_q(n)$ and integers $a'_\bx$, by setting $t=4p\cL$ to account for the deletion of length at most $2p\cL$. This time, for all $\bx\in\ir_q(n)$, $\log a'_\bx\le {\log\|B^{\leq p,\leq 2 p\cL}_{\ir}(\bx)\|+o(\log n)}$. The labeling function $f'$ and integers $a'_\bx$ are used in Construction~\ref{const:code}.

% ////////////////////////////////////////////////////////////////////////////////////////////////

%%%%%%%%%%%%%%%%%%%%%%%%%%%%%%%%%%%%%%%%%%%%%%%

\subsection{The redundancy of the error-correcting codes} \label{subsec:codes_and_redundancy}

In this section, we study the rate and the redundancy of the codes proposed in Constructions~\ref{const:side-channel} and~\ref{const:code} and compare these to those of the state-of-the-art short-duplication-correcting code given in~\cite{jain2017duplication}, which has the highest known asymptotic rate. For an alphabet of size $q$, the asymptotic rate of this code {for short duplications} is $\log \lambda$, where $\lambda$ is the largest positive real root of $x^3 - (q-2)x^2 - (q-3)x - (q-2) = 0$~\cite{chee2020efficient}.

The following lemma shows that the code proposed in~\cite{jain2017duplication} essentially has size $\ir_q(N)$, where $N$ is the length of the code, a fact that will be helpful for comparing the redundancies of the codes proposed here with this baseline.
\begin{lemma}\label{lem:dup_codes_len_n} Let $\codeJain_N$ be the code of length $N$ over alphabet $\Sigma_q$ introduced by~\cite{jain2017duplication} for correcting any number of duplication errors. For $q\ge 4$,
\begin{align*}
    %2\Vert\ir(n)\Vert &\leq \sum_{i=1}^n \Vert\ir_q(i)\Vert\leq n \Vert\ir(n)\Vert, & q=3\\
    %\frac{q-1}{q-2} \Vert\ir_q(n)\Vert&\leq
    \Vert\ir_q(N)\Vert\le \Vert\codeJain_N\Vert\leq \frac{q-2}{q-3} \Vert\ir_q(N)\Vert.
\end{align*}
\end{lemma}

%\begin{lemma}    \label{lem:dup_codes_len_n}
%Given $q\geq 3$, define $\cC^1=\ir(n)\subseteq \Sigma_{q}^n$ denote an error-correcting code for short duplications. Then the code $\cC^1$ achieves the same asymptotic code rate as the code $\cC^2$ in \cite[Construction~C]{jain2017duplication}, the best known code for short duplications. Furthermore, compared to the code $\cC^2$, the extra redundancy of the code $\cC^1$ is upper bounded by $\log \frac{q-2}{q-3}$ bits when $q\geq 4$ and $\log n$ bits when $q=3$, respectively.
%\end{lemma}

\begin{IEEEproof}
As shown in~\cite{jain2017duplication},  $\Vert\codeJain_N\Vert=\sum_{i=1}^N \Vert\ir_q(i)\Vert$. %Each irreducible string can be represented by a path in a subgraph of the De Bruijn graph~\cite{jain2017duplication}, as described in Appendix~\ref{Appdix:lemma:length_buffer}.
Based on \cite[Lemma~14]{Tang2021ECC_Edit}, given $\bu\in \ir_q(N-1)$, there are at least $q-2$ choices for $a\in \Sigma_q$ such that $\bx=\bu a\in \ir_q(N)$. Thus,  $(q-2)\Vert\ir_q(N-1)\Vert \leq \Vert\ir_q(N)\Vert$ and, consequently, $ \Vert \ir_q(N-i) \Vert \leq \frac{\Vert \ir_q(N) \Vert}{(q-2)^i}$. Then we have
\begin{align*}
    \frac{\sum_{i=1}^N \Vert\ir_q(i)\Vert}{\Vert\ir_q(N)\Vert} &\leq  \sum_{j=0}^{N-1} \frac{1}{(q-2)^j} \leq \frac{q-2}{q-3}.   %\qquad\IEEEQEDhere
\end{align*}
\end{IEEEproof}

%\frac{q-1}{q-2}(1-\frac{1}{(q-1)^n})\Vert\ir_q(n)\Vert=\Vert\ir_q(n)\Vert \sum_{j=0}^{n-1} \frac{1}{(q-1)^j}\leq \sum_{i=1}^n \Vert\ir_q(i)\Vert=\sum_{j=1}^n \Vert\ir_q(n-j)\Vert \leq \Vert\ir_q(n)\Vert \sum_{j=0}^{n-1} \frac{1}{(q-2)^j}=\frac{q-2}{q-3}(1-\frac{1}{(q-2)^n})\Vert\ir_q(n)\Vert$.
%Based on \cite[Theorem~27]{jain2017duplication}, the size of the existing code is $ M=\sum_{i=1}^n \Vert\ir(i)\Vert$. If $q=3$, we have $2\Vert\ir(n)\Vert \leq M\leq n \Vert\ir(n)\Vert $. If $q\geq 4$, we have $\frac{q-1}{q-2} \Vert\ir(n)\Vert\leq M\leq \frac{q-2}{q-3} \Vert\ir(n)\Vert$. \begin{align*}
   % \sum_{i=1}^n \Vert\ir_q(i)\Vert &\geq \Vert\ir_q(n)\Vert \sum_{j=0}^{n-1} \frac{1}{(q-1)^j} \\ &\geq \frac{q-1}{q-2}(1-\frac{1}{(q-1)^n})\Vert\ir_q(n)\Vert
%\end{align*}
%and

We now compare the redundancy of the code $\codeA$ of Construction~\ref{const:side-channel} with the best known code $\codeJain$ for correcting only duplications. The length $N$ of $\codeA_n$ is $N=n+|\br|$, where
\begin{equation}\label{eq:r-CA}
|\br| = 2\log_q a_\bx\le 2{\log_q\|B^{\leq p}_{\ir}(\bx)\|+o(\log_q n)}\le 4 p \log_q n+o(\log_q n)
\end{equation}
symbols. Hence, $N = n + 4 p \log_q n+o(\log_q n)$. Then, the difference in redundancies between $\codeA_n$ and $\codeJain_N$, both of length $N$, is
\begin{align}\label{eq:extra}
    \log_q \|\codeJain_N\|-\log_q \Vert \codeA_n\Vert &= \log_q \frac{\Vert \ir_q(N)\Vert}{\Vert \ir_q(n)\Vert} +O(1)\\
    &\le \log_q q^{N-n}+O(1)\\
    &\le 4p\log_q n + o(\log_q n),
\end{align}
%\rcomment{should be $\log_q$.}\\
where the equality follows from Lemma~\ref{lem:dup_codes_len_n} and the first inequality from the fact that $\|\ir_q(i+1)\|\le q \|\ir_q(i)\|$. Noting that $\log_q n=\log_q N+o(\log_q N)$ yields the following theorem.

\begin{theorem}
For constants $q\ge4$ and $p$, the redundancy of the code $\codeA_n$ of length $N$ is larger than the redundancy of the code $\codeJain_N$ of the same length by at most $4\log_q N + o(\log_q N)$.
\end{theorem}

We now turn our attention to comparing the redundancy of $\codeB_n$ of length $N$ with $\codeJain_N$. Here, $N-n = |\br|+O(1)=|\cE_1(a'_\bx, f'(\bx) \bmod a'_\bx)|{+O(1)}$. Similar to~\eqref{eq:extra}, the extra redundancy is then $|\br|+O(1)$, which through $a'_\bx$ depends on $\|B^{\leq p,\leq 2 p\cL}_{\ir}(\bx)\|$,  investigated in the next lemma. The proof of the lemma is in Appendix~\ref{sec:prefix_B}.%The next two lemmas investigate $\|\bb_\bx\|$ and $\|\br\|$.

\begin{restatable}{lemma}{prefixB}
\label{lem:prefix_B}
For $\bx\in \ir_q(n)$ with $q\geq 3$, %the size of $B^{\leq p,\leq 2 p\cL}_{\ir}(\bx)$ satisfies
\begin{align*}
  \Vert B^{\leq p,\leq 2 p\cL}_{\ir}(\bx)\Vert %&\leq  (2p\cL+1)(2p\cL q^{2p\cL}+1) (968 q (n+p\cL)+1)^{2p}\\
   \le q^{4p\cL}(n+p\cL)^{2p}.%(2p\cL+1)^2 (968q+1)^{2p}
\end{align*}
\end{restatable}

\begin{lemma}\label{lem:redundancy}
For constants $q\geq 4$ and $p$, and $\bx\in \ir_q(n)$, the length $|\br|$ of $\br = \cE_1(a'_\bx, f'(\bx) \bmod a'_\bx )$ satisfies
\[|\br|\leq 8p\log_q n+o(\log_q n).\]
\end{lemma}

\begin{IEEEproof}
From the previous subsection, assuming $p$ is a constant, we have that $\log a'_\bx \le \log\|B^{\leq p,\leq 2 p\cL}_{\ir}(\bx)\|+o(\log n)\le 2p\log n + o(\log n)$. Since $(f'(\bx)\bmod a'_\bx)\le a'_\bx$, we need $ 4p\log n+o(\log n)$ bits to represent the pair $(a'_\bx, f'(\bx) \bmod a'_\bx )$. Then, by Lemma~\ref{lem:last_paper}, $|\cE_1(a'_\bx, f'(\bx) \bmod a'_\bx ))|\le 4p\log n (1+o(1))/\log(q-2)$. The lemma follows from  $\frac{\log q}{\log (q-2)}\le 2$ for $q\ge 4$.
%Since the code in~\cite[Construction~10]{tang2021error_atmost_p} has a rate $R(\cC_{E})\geq \log(q-2)(1-o(1))$, the length of $\bsigma\cE_1(a', f'(\bx) \bmod a' )$ satisfies $L=L_r/R(\cC_{E})+l\leq (1+o(1)) L_r \log_{q-2}2+l=(1+o(1))4p\log_{q-2}2 \log(n+p\cL)+o(\log n)$, where $l$ is a constant denoting the length of $\bsigma$.
\end{IEEEproof}

%\bcomment{
%Given a code $\cC\subseteq \Sigma^{n}_q$, the code rate and the asymptotic code rate of the code $\cC$ are defined as $R(\cC_n)=\frac{1}{n}\log \Vert \cC \Vert$ and $R(\cC)=\lim_{n\to \infty}\frac{1}{n}\log \Vert \cC \Vert$, respectively. %Furthermore, the redundancy of the code is $r(\cC_n)=1-R(\cC_n)$. Then we have the following lemma.
Using Lemma~\ref{lem:redundancy}, the next theorem gives the extra redundancy of correcting $p$ substitutions compared to~\cite{jain2017duplication} and shows that there is no relative asymptotic rate penalty.%, in contrast to prior work~\cite{tang2021error_atmost_p} which also corrects duplications and $p$ substitutions.
\begin{theorem}
    For constants $q\ge4$ and $p$, the redundancy of the code $\codeB_n$ of length $N$ is larger than the redundancy of the code $\codeJain_N$ of the same length by at most $8\log_q N + o(\log_q N)$. The codes have the same asymptotic rate, which,  for $q=4$, equals $\log 2.6590$.
\end{theorem}

%\rcomment{Update here!}
%%%%%%%%%%%%%%%%%%%%%%%%%%%%%%%%%%%%%%%%%%%%
%%%%%%%%%%%%%%%%%%%%%%%%%%%%%%%%%%%%%%%%%%%%%

%%%%%%%%%%%%%%%%%%%%%%%%%%%%%%%%%%%%%%%%%%%%%%%%%%%
\iffalse% \\\\\\\\\\\\\\\\\\\\\\\\\\\\\\\\\\\\\\\\\\\\\\\\
\subsection{The encoding process}
Based on the analysis in this section above, we present the encoding process and the error-correcting codes:
\begin{itemize}
      \item Given $q\geq 4$ and $\bsigma\in \ir(5)$, obtain $c_q$.
    \item For an arbitrary irreducible string $\bx\in \ir(n)$, find a buffer $\bs_b\in \ir(c_q)$ such that $\bx\bs_b\bsigma\in \ir(n+c_q+5)$.
    \item Based on $\bx\in \ir(n)$, derive the labeling function $f'(\bx)$ and $a' \leq  2\log  ||B_{\ir}^{\leq p,\leq 1}(\bx)||+o(\log n)$, where $a'$ is identified by a brute force search. %\textcolor{red}{Maybe be more specific and say identify $a'$ by brute force. The labeling function is already known.} \ggcomment{Yuanyuan: I revised it.}
    \item Given $a'$ and $f'(\bx)$, we generate $\bsigma\cE_1(a', f'(\bx) \!\!\mod a' )\in \ir(L)$.
    \item By concatenating $\bx$, $\bs_b$, and $\bsigma\cE_1(a', f'(\bx) \!\!\mod a' )$, we generate the codeword
    \[\bz=(\bx, \bs_b, \bsigma\cE_1(a', f'(\bx) \!\!\mod a' ))\]
    with the overall redundancy $L+c_q \simeq 4p\log_{q-2}2 \log{(n+p\cL)}$ if $n\to \infty$ and $p$ is a constant with respect to $n$.
\end{itemize}
Note that the decoding process is presented in Subsection~\ref{subsec:decoding}.
\fi% ////////////////////////////////////////////////////////////////////////////////////////////////

%%%%%%%%%%%%%%%%%%%%%%%%%%%%%%%%%%%%%%%%%%%%%%%%%%%%%%%%%

\subsection{Time complexity of encoding and decoding}\label{secIV:time_complexity}

%%%%%%%%%%%%%%%
% Short version of time complexity

%The encoding process relies on determining a value for $a'$ among at most $2^{\log \|B(\bx)\|+o(\log n)}$ satisfying the condition discussed in Subsection~\ref{subsec:labeling_function}. This step has complexity $O(n^{4p+1})$ in $n$, making the total complexity of encoding polynomial in $n$. Decoding requires deduplication, which is linear in the length of the retrieved sequence, and a brute-force search among all inputs that can lead to the same output $(n-p\cL)$-prefix of the root of the retrieved sequence, which is polynomial in $n$. Hence, decoding is polynomial in the length of the retrieved sequence.

%\rcomment{Yuanyuan: Is it necessary to present a discussion about time complexity in detail?}

Suppose $q\ge 4$ is a constant. The total time complexities of both the encoding and decoding processes are polynomial in the lengths of the stored and retrieved sequences, respectively%,  mainly resulting from the brute force search when $p$ is a constant with respect to $n$
. The encoding process consists of four main parts:
\begin{enumerate}
    \item Generating $\bx\in \ir_q(n)$ by the state-splitting algorithm, which has polynomial-time complexity~\cite{chee2020efficient}. %Given a constant $q>3$ and $k=3$, the time complexity of the state-splitting encoder is almost linear of $n$~\cite{chee2020efficient}.
    \item  Determining $\bb_\bx$ such that $\bx \bb_\bx \bsigma\in \ir_q(*)$, which has constant time complexity as the relevant subgraph of the De Bruijn graph (see Appendix~\ref{Appdix:lemma:length_buffer}) has a constant size (no more than $q^5$ vertices).% Then the time complexity to find $\bb_\bx$ is constant.

    \item  Determining $a'_\bx$ and $f'(\bx) \bmod a'_\bx$. This is done in three steps, with polynomial time complexity. i) Given $\bx \in \ir_q(n)$, we find the elements of a set $\hat B \supseteq B_{\ir}^{\leq p,\leq 2 p\cL}(\bx)$ whose size satisfies the upper bound given in Lemma~\ref{lem:prefix_B}. Specifically, given $\bx$ we find all sequences that can be obtained from it through $\le p$ \emph{short substring substitutions}, one deletion of a suffix of length $\leq 2p\cL$, one insertion of a suffix of length $\leq 2p\cL$, and another $\le p$ {short substring substitutions}, where in each short substring substitution step, we replace a substring $abcde\in \ir_q(5)$ by another irreducible substring from the set $ R(\ddes{1}({abcde}))$ and then deduplicate all copies. The total time complexity of this step is $O(n^{2p})$ as each element of $\hat B$ is obtained by a bounded number of operations and $\|\hat B\|=O(n^{2p})$.
    ii) Since computing $f'(\cdot)$ from~\cite{sima2020optimal_Systematic} has time complexity $O(n \log n )$, computing it for all elements of $\hat B$ takes $O(n^{3p} \log n)$ steps. iii) Computing the remainder of these values modulo the $\le 2^{\log O(n^{2p})}$ possible values for $a_{\bx}'$ also has polynomial complexity.
    %iii) Third, given $f'(\bx)$ and $B_{\ir}^{\leq p,\leq 2 p\cL}(\bx)$, we determine a value $a'_\bx \leq 2^{\log \|B^{\leq p,\leq 2p\cL}(\bx)\|+o(\log n)}$ with the time complexity $O(n^{4p+1})$.
    \item  Generating $\br = \cE_1(a'_\bx, f'(\bx) \bmod a'_\bx )$ using the encoder $\cE_1$ for the code in  Construction~\ref{con:CMDS_Ecc_for_channels}, which has complexity polynomial in $|\br|$ based on Subsection~\ref{subsec_V:C_E_time_complexity}. Hence, by Lemma~\ref{lem:redundancy}, the complexity is at most polynomial.
    %\item  For iv), suppose $L=|\bsigma\cE_1(a'_\bx, f'(\bx) \bmod a'_\bx )|=O(\log n)$, we generate $\bsigma\cE_1(a'_\bx, f'(\bx) \bmod a'_\bx )$ by four steps with the time complexity $O(\log^2 n)$.  Step 1), by Construction~\ref{con:CMDS_Ecc_for_channels} and Section~\ref{Sec:Aux_ECC}, the binary string $(a'_\bx, f'(\bx) \bmod a'_\bx )$ is divided into $O(L/m)$ length-$m$ binary blocks such that $mT \times \frac{(q-2)^m}{2T}=O(L)$, where $T=3p$ and $m=O(\log L)$. Step 2), after separating all length-$m$ binary blocks into $T$ groups, we generate $T$ MDS codeword such that each codeword has $d_H=4p+1$. The time complexity is $O(L^2)$. Step 3) we construct a mapping table between $\{0,1\}^*$ and $\cB^m_{\bsigma}(j)$ for $j\in [T]$ with size $\leq (q-2)^m/T-1=O(L)=O(\log n)$, where these elements are lexicographically listed. Step 4), we map each binary string to an irreducible string with complexity $O(m)$ by the binary tree search method. Since there are $O(L/m)$  message blocks, the time complexity of mapping all binary strings to irreducible message blocks is $O(\log n)$.
\end{enumerate}
Therefore, when $p$ is a constant, the time complexity of the encoding process is polynomial with respect to $N$ (as well as $n$).

 Decoding requires finding the root of the retrieved word, which is linear in its length; decoding $a'_\bx$ and $f'(\bx)\bmod a'_\bx$, which is polynomial as discussed in Subsection~\ref{subsec_V:C_E_time_complexity}; and  determining $\bx$ through a brute-force search among all inputs that can lead to the same $(n-p\cL)$-prefix of the root of the retrieved sequence. Similar to the discussion about finding $\hat B$ above, the brute-force search is polynomial in $n$. Hence, decoding is polynomial in the length of the retrieved sequence.

% The time complexity of the decoding algorithm

\iffalse %{{{{{{{{{{{{{{{{{{{{{{{{{{{{{{{{{{{{{{{{{{{{
The decoding process can be considered as three steps: i) get the root of the received sequence by deduplications; ii) recover $(a'_\bx, f'(\bx) \mod a'_\bx)$ by the MDS code; and iii) decode $\bx$ by the brute-force search.
\begin{itemize}
    \item For i), the time complexity for deduplications is  linear in the length of the retrieved sequence.
    \item For ii), the decoding process to recover $(a'_\bx,f'(\bx) \mod a'_\bx)$ consists of identifying message groups, mapping irreducible message blocks to binary sequences, and decode $(a'_\bx,f'(\bx) \mod a'_\bx)$ by the MDS decoder. Based on the forth part of the encoding process, the time complexity of the MDS decoder is also upper bounded by $O(\log^2n)$.
    \item  For iii), the decoding of $\bx$ by performing a brute force search has the time complexity $O(n^{p+3})$. Given an output $\bs\in D^{*, \leq p,\leq 2p\cL}(\bx)$, the number of confusable inputs $\by\in  B_{\ir}^{\leq p,\leq 2 p\cL}(\bx)$ is bounded by $O(n^{p+1})$.  By a brute force search, we compute all $f'(\by)$ for all $\by$ derived from $\bs$,
    leading to the time complexity upper bounded by $O((n)^{p+1} \cdot n\log n)=O(n^{p+3})$.
\end{itemize}
 Therefore, the total time complexity for the decoding process polynomial.
\fi %}}}}}}}}}}}}}}}}}}}}}}}}}}}}}}}}}}}}}}}}}}}}}}}}}}}}}}

%%%%%%%%%%%%%%%%%%%%%%%%%%%%%%%%%%%%%%%%%%%%%%%%%%%%%%%%%%%%%%%%%%%%%%%%%%%%%%%%%%%%%%%%%%%%%%%%%%%%%%%%%%%%%%%%%%%%%%%%%%%%%%%%%%%
%%%%%%%%%%%%%%%%%%%%%%%%%%%%%%%%%%%%%%%%%%%%%%%%%%%%%%%%%%%%%%%%%%%%%%%%%%%%%%%%%%%%%%%%%%%%%%%%%%%%%%%%%%%%%%%%%%%%%%%%%%%%%%%%%%%%

%%%%%%%%%%%%%%%%%%%%%%%%%%%%%%%%%%%%%%%%%%%%%%%%%%%%%%%%
%%%%  MDS code for short TDs and substitutions
%%%%%%%%%%%%%%%%%%%%%%%%%%%%%%%%%%%%%%%%%%%%%%%%%%%%%%

\section{Auxiliary high-redundancy error-correcting codes}\label{Sec:Aux_ECC}

Based on lemma~\ref{lem:last_paper} in Section~\ref{sec:ECC}, the error-correcting codes for short duplications and at most $p$ substitutions with low redundancy rely on an error-correcting code to protect the syndrome information $(a'_\bx, f'(\bx) \bmod a'_\bx)$, where $(a'_\bx, f'(\bx) \bmod a'_\bx)$ is considered as a binary sequence. Therefore, this section focuses on constructing error-correcting codes that can protect this information from short duplications and at most $p$ substitutions. We will also present the rate of the proposed codes in Section~\ref{sec:code_rate}, followed by the proof of Lemma~\ref{lem:last_paper} used in the previous section.% that can protect the syndrome information $(a'_\bx, f'(\bx) \bmod a'_\bx)$.

While in the previous section, we used syndrome compression with a labeling function designed to handle indel errors, in this section, the errors are viewed as substring edits in irreducible sequences, as described in~Theorem~\ref{Theo:root_L_17}. An example for Theorem~\ref{Theo:root_L_17} is given in Appendix~\ref{app:example}.

%%%%%%%%%%%%%%%%%%%%%%%%%%%%%%%%%%%%%%%%%%%%%%%%%%
%%%%%%%%%%%%%%%%%%%%%%%%%%%%%%%%%%%%%%%%%%%%%%%%%%%%

\subsection{Code construction}\label{subsec:code_construction}

To construct codes correcting at most $p$ $\cL$-substring edits in irreducible sequences, similar to \cite{Tang2021ECC_Edit}, we divide the codewords into message blocks, separated by markers, while maintaining irreducibility, such that an $\cL$-substring edit only affects a limited number of message blocks. In the case of $p=1$ studied in~\cite{Tang2021ECC_Edit}, it was shown that if the markers appear in the correct positions in the retrieved word, then at most two of the message blocks are substituted. For $p>1$ however, even if all markers are in the correct positions, it is not guaranteed that a limited number of message blocks are substituted, making it challenging to correct more than one error.

We start by recalling an auxiliary construction from~\cite{Tang2021ECC_Edit}.

\begin{construction}\label{cons:blocks_separated}
\cite[Construction 6]{Tang2021ECC_Edit} Let $l,m,\nblk$ be positive integers with $m>l\ge5$ and $\bsigma\in \ir_q(l)$. Also, let $\cB_{\bsigma}^m$ denote the set of sequences $B$ of length $m$ such that ${\bsigma}B{\bsigma}$ is irreducible and has exactly two occurrences of ${\bsigma}$. Define
%$\cC_{\bsigma}$ is given as
\[
\cC_{\bsigma} = \{B_1\bsigma B_2 \bsigma \dotsm \bsigma B_{\nblk}: B_i\in\cB_{\bsigma}^m\}.
\]
\end{construction}

The irreducibility of $\bsigma B_i\bsigma$ ensures that the codewords are irreducible.

We denote the output of the channel by $\by$. Define a \textit{block} in $\by$ as a maximal substring that does not overlap with any $\bsigma$. Furthermore, define an \emph{$m$-block} in $\by$ as a block of length $m$. Note that $m$-blocks can be either message blocks in $\bx$ or new blocks created by substring edits.
%Let an $m$-block in $\by$ be  \emph{error-free} if it is derived from a message block in $\bx$ with no errors.

Having divided each codeword into $\nblk$ message blocks and $\nblk-1$ separators, %it is significant to obtain effects of $\leq p$ substring edits over $N$ message blocks.
we study in the next lemma how message blocks are affected by the errors.

%%%%%%%%%%%%%%%%%%%%%%%%%%% Comment start
%%%%%%%%%%%%%%%% Comments end

%After discussing the case with only one $p\cL$-substring edit, i.e., $\bbeta_i\to \bbeta'_i$,  we can derive the number of $m$-blocks in $\by$ if $z$ $p\cL$-substring edits occur in $\by$ with $z\leq p$. (Recall that the sum of affected length of $z$ $p\cL$-substring edits is also bounded by $\leq p\cL$.)

%\rcomment{Some of the following results only deal with substring edits and not about symbol substitution. We should state them by saying $\by$ is obtained from $\bx$ through $\le p$ substring edits rather than $\by\in R(D(x))$. [Do not delete this comment]}

\begin{lemma}\label{lem:blocks_receiver}
Let $\bx\in \cC_{\bsigma}$,
%$\by\in R(\ddes{\leq p}(\bx))$,
 $m>\cL$, and $\by$ be generated from $\bx$ through at most $p$ $\cL$-substring edits.
Then there are less than $(\nblk+p)$ $m$-blocks in $\by$. Furthermore, there are at least $\nblk-2p$ error-free $m$-blocks in $\by$ which appear in $\bx$ in the same order. More precisely, there are blocks $B_{i_1},B_{i_2},\dotsc,B_{i_k}$ in $\by$, where $k\ge \nblk-2p$, each $B_{i_j}$ is a message block in $\bx$, and any two blocks $B_{i_j}$ and $B_{i_{j'}}$ have the same relative order of appearance in $\bx$ and in $\by$.
\end{lemma}

\begin{IEEEproof}
First suppose $\by$ has $\ge(\nblk+p)$ message blocks. This implies that the length of $\by$ is at least $(\nblk+p)m+(\nblk+p-1)l$, which is larger than the length of $\bx$ by $pm+(p-1)l$. But this is not possible as $m>\cL$ and the total length of inserted substrings is at most $p\cL$.

Furthermore, if $m>\cL$, each $\cL$-substring edit alters i) a message block in $\bx$, ii) a message block and a marker $\bsigma$, or iii) two message blocks and the marker between them. Hence at least $\nblk-2p$ message blocks of $\bx$ appear in $\by$ without being changed.
\end{IEEEproof}

%Lemma~\ref{lem:blocks_receiver} shows the number of error-free $m$-blocks (also message blocks) that can be detected in $\by$.
If the positions of the error-free $m$-blocks described in Lemma~\ref{lem:blocks_receiver} in $\by$ were known, a Reed-Solomon (RS) code of length $\nblk$ and dimension $\nblk-2p$ could be used to recover codewords in $\cC_{\bsigma}$. This however is not the case since the blocks can be shifted by substring edits. %However, given $\bx\in \cC_{\sigma}$ passing through the $p$-concatenating channel, positions of error-free $m$-blocks are ambiguous.
In order to determine the  positions of the error-free $m$-blocks, we introduce another auxiliary construction based on Construction~\ref{cons:blocks_separated} by combining message blocks into \emph{message groups}, where the message blocks in each group have different ``colors''. %The message blocks in each group have  and adding \emph{colors} to message blocks, where \bcomment{colors} in message blocks (also $m$-blocks) can be recognized in $\by$.

%To consider message blocks independently, the separator $\bsigma\in \ir(l)$ should eliminate the interaction of adjacent message blocks in $\bx$. Based on \cite[Theorem 5]{tang2020error}, given $l\geq 5$, $\bx\in \ir(N(m+l)-l)$ is satisfied if $\bsigma B_i \bsigma \in \ir(m+2l)$.

\begin{construction}\label{const:message_group}
For an integer $T$, we partition $\cB_{\bsigma}^m$  into $T$ parts $\cB_{\bsigma}^m(j), j\in [T]$. The elements of $\cB^m_{\bsigma}(j)$ are said to have \emph{color} $j$. Let $\nblk$ be a positive integer that is divisible by $T$. We define the code
\begin{equation*}
    \cC_{(\bsigma,T)}=\left\{ B_1\bsigma B_2\bsigma \dotsm \bsigma B_{\nblk} \in \cC_{\bsigma}: B_{i}\in\cB_{\bsigma}^m(i \shiftedMod T)
    %\\\{B_{j}, B_{j+T},\dotsc,B_{j+(\hat N-1)T} \} \subseteq \cB_{\bsigma}^m(j), j\in [T]
    \right\},
\end{equation*}
where $\cC_{\bsigma}$ has parameters $m,l$ with $m>\cL$ and $m>l\geq 5$.
%Furthermore, we define $\cB_{\bsigma}^m (0)$ as $\cB_{\bsigma}^m(T)$.

We divide the message blocks $B_1,\dotsc,B_\nblk$ in each $\bx\in \cC_{(\bsigma,T)}$ into $\hat N=\nblk/T$ \emph{message groups}, where the $k$-th message group is %the set of $T$ message blocks in the form $\bsigma B_{(k-1)T+1}\bsigma \dotsc\bsigma B_{kT-1} \bsigma B_{kT} \bsigma$ with \bcomment{colors} $\{1,2,\dotsc, T\}$, i.e.,
$S_{k}=(B_{(k-1)T+1},\dotsc,B_{kT-1},B_{kT})$. Note that the message blocks in each message group have colors $1,2,\dotsc,T$ in order. %Then $\bx$ has $\hat N$ message groups.
\end{construction}

For example, if $\nblk=12,T=3,\hat N=4$, then in a codeword  \[\bx={\color{red} B_1}\bsigma {\color{Brown} B_2}\bsigma {\color{blue}B_3}\bsigma {\color{red} B_4}\bsigma {\color{Brown}B_5}\bsigma {\color{blue}B_6}\bsigma\dotsm\bsigma {\color{red} B_{10}}\bsigma {\color{Brown}B_{11}}\bsigma {\color{blue}B_{12}},\] the first group is $(B_1,B_2,B_3)$ and the second group is $(B_4,B_5,B_6)$. Furthermore, message blocks in both groups have {colors} $({\color{red}1},{\color{Brown}2},{\color{blue}3})$. The colors in the message group will help us identify the true positions of the message blocks.

\begin{definition}\label{Def:T-groups}
For $\bx\in \cC_{(\bsigma, T)}$ and $\by$ derived from $\bx$ through at most $ p$ $\cL$-substring edits,
%$\by\in R(\ddes{\leq p}(\bx))$,
let the $i$-th $m$-block in $\by$ be denoted by $B_i'$. A \emph{$T$-group} in $\by$ is a substring $B'_{k+1} \bsigma B'_{k+2}  \dotsm \bsigma B'_{k+T}$ such that the $m$-block $B'_{k+j}$ has color $j$.
\end{definition}

%\bcomment{\begin{definition}\label{Def:T-groups_cyclic}For $\bx\in \cC_{(\bsigma, T)}$ and $\by$ derived from $\bx$ through at most $ p$ $\cL$-substring edits,
%$\by\in R(\ddes{\leq p}(\bx))$,let the $i$-th $m$-block in $\by$ be denoted by $B_i'$. A  \emph{$\eta$-cyclic $T$-group} in $\by$ is a substring $B'_{k+1} \bsigma B'_{k+2}  \dotsm \bsigma B'_{k+T}$ such that the $T$ $m$-blocks have colors $\{\eta,\eta+1,\dotsc, T,1,2, \dotsc, \eta-1\}$ in order, where $\eta\in[T]$. \end{definition} }

The next lemma characterizes how error-free message groups (those that do not suffer any substring edits but may be shifted) appear in $\by$. %The next lemma provides a lower bound on the number of message groups not affected by substring edit errors and thus appearing as $T$-blocks in $\by$. Note that even though these $T$-groups do not suffer any substring edits, they may still be shifted because of edits in other positions in $\bx$.

%Similar to $m$-blocks, let a $T$-group be \emph{error-free} if it is directly a message group in $\bx$. Based on the definition, the $T$-groups in $\by$ have the following properties.

\begin{lemma}\label{lem:T-groups}
Suppose $\bx\in \cC_{(\bsigma, T)}$ and let $\by$ be obtained from $\bx$ through at most $ p$ $\cL$-substring edits.
%$\by\in R(\ddes{\leq p}(\bx))$. % be obtained from $\bx$ through  $\leq p$ equivalent $\cL$-substring edits.
For $r\in[\hat N]$, if the $r$-th message group in $\bx$ is not affected by any substring edit errors, then it will appear as a $T$-group after $b$ $m$-blocks in $\by$, where $b\in[(r-1)T-2p,(r-1)T+p-1]$.
\end{lemma}

\begin{IEEEproof}
%Given $\bx\in \cC_{(\bsigma, T)}$ and let $\by\in R(\ddes{\leq p}(\bx))$. Based on Theorem~\ref{Them:errors_pIS_roots}, $\by$ can be derived from $\bx$ after $\leq p$ equivalent $\cL$-substring edit errors. Since $\bx\in \cC_{(\bsigma, T)}$, then $\bx$ has $\hat N$ message groups. According to Lemma~\ref{lem:blocks_receiver}, given $m>\cL$, a $\cL$-string edit at most affect two message blocks and two message groups. In the worst case, for the channel with $\leq p$ $\cL$-string edits, at least $(\hat N-2p)$ message groups in $\bx$ are not affected by any errors. Therefore, at least $(N-2p)$ $T$-groups are error-free in $\by$.
Since $m>\cL$, each $\cL$-substring edit can affect at most two message blocks and thus at most two message groups. Hence, there are at least $\hat N-2p$ message groups that do not suffer any substring edits.

Let the $r$-th message group $S_{r}$ in $\bx$ be free of substring edits. Given that the colors of its message blocks are not altered, it  will appear as a $T$-group in $\by$. Since each substring edit alters at most two message blocks, among the $(r-1)T$ message blocks appearing before $S_{r}$ in $\bx$, at most $2p$ do not appear in $\by$. Furthermore, the substring edits add at most $p\cL$ to the length of $\bx$. Since $m>\cL$, this means that at most $p-1$ new $m$-blocks are created in $\by$. Hence, $b\in[(r-1)T-2p,(r-1)T+p-1]$.
\end{IEEEproof}

The previous lemma guarantees the presence of error-free, but possibly shifted, $T$-groups, and provides bounds on their position in $\by$. In the following theorem, we use these facts to show that these $T$-groups can be synchronized and the errors can be localized.

\begin{theorem}\label{Theo:pIS_channel_model}
Let $\cC_{(\bsigma,T)}$ be a code in Construction~\ref{const:message_group} and suppose $T\geq 3p$ and $\hat N\geq 4p+1$. There is a decoder $\dec$ such that, for any $\bx\in \cC_{(\bsigma,T)}$ and $\by$ derived from $\bx$ through at most $p$ $\cL$-substring edits,
%$\by=R(\ddes{\leq p}(\bx))$ relative to $\bx$,
$\bv=\dec(\by)$ suffers at most $t$ substitutions and $e$ erasures of message groups, where $t+e \leq  2p$.
%$2t$ substitutions and $4e$ erasures of message groups, where $t+e=z$ with $z\leq p$.
\end{theorem}

\begin{IEEEproof}
%\bcomment{
We start by identifying all $T$-groups in $\by$. Note that no two $T$-groups can overlap. Let $\bv=(S_1',\dotsc,S_{\hat N}')$ be the decoded vector, where $S'_r$ is the decoded version of the message group $S_r$, determined as follows.

For $r=1,\dotsc,\hat N$:
\begin{enumerate}
    \item If there exists a $T$-group $\cT$ appearing after $b$ message blocks such that $b\in [(r-1)T-2p,(r-1)T+p-1]$, then let $S_r' = \cT$.
    \item If such a $T$-group does not exist, let $S_r'=\Lambda$, denoting an erasure.
\end{enumerate}
We note that for each $r$, at most one $T$-group may satisfy the condition in~1). If two such $T$-groups exist appearing after $b$ and $b'$ message blocks, we must have $|b-b'|\ge T$ and $b,b'\in[(r-1)T-2p,(r-1)T+p-1]$, implying $3p-1\ge T$, which contradicts the assumption on $T$.

If a message group $S_r$ is not subject to a substring edit, then by Lemma~\ref{lem:T-groups}, we have $S_r'=S_r$. Otherwise, we may have a substitution of that message group, i.e., $S_r'\neq S_r$, or an erasure, $S_r'=\Lambda$. Since each substring edit may affect at most $2$ message groups, the total number of substitutions and erasures is no more than $2p$.
%}
\end{IEEEproof}

%Given a codeword $\bx\in \cC_{(\bsigma, T)}$ and $\bv=\dec(\by)$ for $\by\in R(\ddes{\leq p}(\bx))$, according to Theorem~\ref{Theo:pIS_channel_model}, at least $\hat N-2p$ message blocks in $\{B_{j}, \dotsc, B_{j+kT}, \dotsc, B_{j+(\hat N-1)T} \}\subseteq \cB_{\bsigma}^m(j), j\in [T],$ are error-free with accurate positions in $\bv$. Then $T$ MDS codes over $\bx\in \cC_{(\bsigma,T)}$ can decode $\bx$.

We now construct an MDS code that can correct the output of the decoder of Theorem~\ref{Theo:pIS_channel_model}.

\begin{construction}\label{con:CMDS_Ecc_for_channels}
 %\cite[Construction~11]{tang2021error_atmost_p}
 Let $\cC_{(\bsigma,T)}$ be the code in Construction~\ref{const:message_group} with parameters $l,m, T,\hat N$ satisfying $m>\cL, m>l\geq 5, T\geq 3p$, and $\hat N\ge 4p+1$.
 Furthermore, assume $|\cB_{\bsigma}^m(j)|\geq \hat N+1$ for $j\in [T]$. Finally, let $\gamma$ be a positive integer such that $2^{\gamma}\le \hat N+1$ and $\zeta_j:\bbF_{2^{\gamma}} \to \cB_{\bsigma}^m(j)$ be an injective mapping for $j\in [T]$.
We define $\cC_{E}$ as
\begin{equation*}
\begin{split}
        \cC_{E}= \{&\zeta_1(c^1_1){\bsigma}\dotsm \bsigma \zeta_j(c^j_1){\bsigma}\dotsm{\bsigma}\zeta_T(c^T_1){\bsigma}  \\
       & \zeta_1(c^1_2){\bsigma}\dotsm \bsigma \zeta_j(c^j_2){\bsigma}\dotsm{\bsigma}\zeta_T(c^T_2){\bsigma}\dotsm \\&
       \zeta_1(c^1_{\hat N}){\bsigma}\dotsm \bsigma \zeta_j(c^j_{\hat N}){\bsigma}\dotsm{\bsigma}\zeta_T(c^T_{\hat N})\!:\\
       &\!\{\bc^{j},j\in [T]\}\subseteq \mds(\hat N, \hat N-4p,4p+1)\},
\end{split}
\end{equation*}
where $\mds(\hat N, \hat N-4p,4p+1)$ denotes an MDS code over $\bbF_{2^{\gamma}}$ of length $\hat N=2^{\gamma}-1$, dimension $\hat N-4p $, and minimum Hamming distance $d_H=4p+1$, and  $\bc^{j} = (c^j_1, c^j_2, \ldots, c^j_{\hat N})$ is a codeword of the MDS code.
\end{construction}

For each $j$, we also define an inverse $\zeta_j^{-1}$ for $\zeta_j$. For $B\in\cB_{\bsigma}^m(j)$, if $\beta\in\mathbb F_{2^\gamma}$ such that $\zeta_j(\beta)=B$ exists, then let $\zeta_j^{-1}(B)=\beta$. Otherwise, let $\zeta_j^{-1}(B)=0$.

\begin{theorem}\label{Them:ECC_for_channels}
The error-correcting codes $\cC_{E}$ in Construction~\ref{con:CMDS_Ecc_for_channels} can correct any number of short duplications and at most $p$ symbol substitutions.
\end{theorem}

\begin{IEEEproof}
Given a codeword $\bx \in \cC_{E}$, let $\bx''\in \ddes{\leq p}(\bx)$  and let $\by=R(\bx'')$. Note that by construction, $\bx$ is irreducible. Thus, by Theorem~\ref{Theo:root_L_17}, $\by$ can be obtained from $\bx$ through at most $p$ $\cL$-substring edits. As $\cC_{E}\subseteq\cC_{(\bsigma,T)}$, based on Theorem~\ref{Theo:pIS_channel_model},  $\bv=\dec(\by)$ suffers at most $t$ substitutions and $e$ erasures of message groups, where $t+e\leq 2p$. Hence, for $j\in[T]$, the blocks $(\zeta_j(c^j_1), \zeta_j(c^j_2), \dotsc, \zeta_j(c^j_{\hat N}))$ suffer at most $2p$ erasures or substitutions. Consequently, if we apply $\zeta^{-1}_j$ to the corresponding retrieved blocks in $\bv$, the codeword $(c_1^j,c_2^j,\dotsc,c_{\hat N}^j)$ also suffers at most $2p$ substitutions or erasures, which can be corrected using the MDS code.
%
%, $j\in [T]$, belongs to $\hat N$ message groups, then at least $\hat N-2p$ message blocks are not affected by any errors. To correct those substitution and erasure errors, the minimum Hamming distance of a code should satisfy $d_H\geq \max\{2t+e\}+1=4p+1$.
%
% Based on Construction~\ref{con:CMDS_Ecc_for_channels}, given a codeword $\bx\in \cC_{E}$, $\{\zeta_j(c^j_1), \zeta_j(c^j_2), \dotsc, \zeta_j(c^j_{\hat N})\}\subseteq \cB_{\bsigma}^m(j)$ denote $\hat N$ message blocks  $\{B_{j},B_{j+T}, \dotsc, B_{j+kT}, \dotsc, B_{j+(\hat N-1)T} \}$. Furthermore, there is a one-to-one mapping between $\{\zeta_j(c^j_1), \zeta_j(c^j_2), \dotsc, \zeta_j(c^j_{\hat N})\}$ and $\{c^j_1, c^j_2, \dotsc, c^j_{\hat N}\}$. Since the codeword $\bc^j=c^j_1 c^j_2 \dotsm c^j_{\hat N}$ belongs to $\mds (\hat N, \hat N-4p,4p+1)$ with $d_H=4p+1$, it can correct $t$ substitutions and $e$ erasures to recover $\bx$ from $\by$.  Hence, it can correct $\leq p$ $\cL$-substring edits, equivalently an arbitrary number of $\leq 3$-TDs and at most $p$ symbol substitutions.
%This completes the proof of Theorem~\ref{Them:ECC_for_channels}.
\end{IEEEproof}

%%%%%%%%%%%%%%%%%%%%%%%%%%%%%%%%%%%%%%%%%%%%%%%%%%%%%%%
\subsection{Code rate}\label{sec:code_rate}
In this subsection, we present choices for the parameters of Construction~\ref{con:CMDS_Ecc_for_channels} and discuss the rate of the resulting code.

Among the $n_E$ symbols of each codeword in Construction~\ref{con:CMDS_Ecc_for_channels},
\(
4pTm+(\hat N T-1)l
\)
symbols belong to MDS parities or markers. We choose $T$ and $l$ to be their smallest possible values and set $T=3p$ and $l=5$.

The construction requires that $\|\cB_{\bsigma}^m(j)\|\ge \hat N+1$ for all $j$. Let $M_{\bsigma}^{(m)}=\|\cB_{\bsigma}^m\|$.
%\rcomment{I did not add parens for $m$ in the notation of $M$. Needs to be added }
Dividing $\cB_{\bsigma}^m$ into parts of nearly equal sizes, we find that each part $\cB_{\bsigma}^m(j)$ has size at least $M_{\bsigma}^{(m)}/T-1$. We then choose $\hat N+1$ as the largest power of two not larger than $M_{\bsigma}^{(m)}/T-1$, ensuring that $\hat N+1\ge M_{\bsigma}^{(m)}/(2T)-(1/2)$.
Assume
\begin{equation}\label{eq:assump-on-M}
    M_{\bsigma}^{(m)}\ge 24p^2+15p.
\end{equation} Then $\hat N+1 \ge M_{\bsigma}^{(m)}/(2T)-(1/2)\ge  4p+2$.

Furthermore, note that $\hat N T(m+5)-5 = n_E$ and thus $\hat N = \frac{n_E+5}{(m+5)(3p)}$. %Then the code size is
The size of the code then becomes
\[\|\cC_{E}\| = (\hat N+1)^{(\hat N-4p)(3p)},\]
and\\
\resizebox{\linewidth}{!}{
\begin{minipage}{\linewidth}
\begin{align}
    \log \|\cC_{E}\|  &\ge \left(\frac{n_E}{m+5}-12p^2\right)\log \left(\frac{M_{\bsigma}^{(m)}}{6p}-\frac12\right)\nonumber\\
    & \ge \left(\frac{n_E}{m+5}-12p^2\right)\left(\log M_{\bsigma}^{(m)}+\log \left(\frac{1}{6p}-\frac1{2M_{\bsigma}^{(m)}}\right)\right)\nonumber\\
     & \ge \left(\frac{n_E}{m+5}-12p^2\right)\left(\log M_{\bsigma}^{(m)}-\log \left(6p+1\right)\right),\label{eq:log-code-size}
  %  \rcomment{&\ge \left(\frac{n}{m+5}-12p^2\right)\log \left(\frac{M_{\bsigma}^{(m)}-3p}{6p}\right)}
\end{align}
\end{minipage}
}
\iffalse% \\\\\\\\\\\\\\\\\\\\\\\\\\\\\\\\\\\\\\\\\\\\\\\\
\begin{align}
    &\log |\cC_{E}|  \ge \left(\frac{n}{m+5}-12p^2\right)\log \left(\frac{M_{\bsigma}^{(m)}}{6p}-\frac12\right)\nonumber\\
    &\quad \ge \left(\frac{n}{m+5}-12p^2\right)\left(\log M_{\bsigma}^{(m)}+\log \left(\frac{1}{6p}-\frac1{2M_{\bsigma}^{(m)}}\right)\right)\nonumber\\
     &\quad \ge \left(\frac{n}{m+5}-12p^2\right)\left(\log M_{\bsigma}^{(m)}-\log \left(6p+1\right)\right),\label{eq:log-code-size}
  %  \rcomment{&\ge \left(\frac{n}{m+5}-12p^2\right)\log \left(\frac{M_{\bsigma}^{(m)}-3p}{6p}\right)}
\end{align}
\fi% ////////////////////////////////////////////////////////////////////////////////////////////////
where in the last step we have used the fact that $ M_{\bsigma}^{(m)}\ge 24p^2+15p.$

It was shown in~\cite{Tang2021ECC_Edit} that $M_{\bsigma}^{(m)}\ge (q-2)^{m-c_q}$ for some $\bsigma$, where $c_q$ is a constant independent of $m$. In particular, $c_3\le 13, c_4\le 7, c_5\le 6$, and $c_q\le 5$ for $q\ge 6$. To satisfy \eqref{eq:assump-on-M}, we need
\begin{equation}\label{eq:assump-on-m}
    m\ge \max\{ \log_{q-2}(24p^2+15p)+c_q,\cL+1\}.
\end{equation}

%Given a code $\cC(n)$ with length $n$ and the code size $|\cC(n)|$, the code rate is defined as $R(\cC(n))=\frac{1}{n}\log|\cC(n)|$.
From~\eqref{eq:log-code-size}, for the rate of $\cC_{E}$,\\
\resizebox{\linewidth}{!}{
\begin{minipage}{\linewidth}
\begin{align*}
    %R(\cC_{E})
    \frac{\log \|\cC_{E}\|}{n_E}
    &\ge \left(\frac{m-c_q}{m+5}-\frac{12p^2m}{n_E}\right)\log(q-2)-\frac{\log(6p+1)}{m+5}\\
    &\ge \left(1-\frac{c_q+5}{m+5}-\frac{12p^2m}{n_E}\right)\log(q-2)-\frac{\log(6p+1)}{m+5},
\end{align*}
\end{minipage}
}
\iffalse% \\\\\\\\\\\\\\\\\\\\\\\\\\\\\\\\\\\\\\\\\\\\\\\\
\begin{align*}
    &R(\cC_{E}) \ge \left(\frac{m-c_q}{m+5}-\frac{12p^2m}{n}\right)\log(q-2)-\frac{\log(6p+1)}{m+5}\\
    &\qquad\ \,\ge \left(1-\frac{c_q+5}{m+5}-\frac{12p^2m}{n}\right)\log(q-2)-\frac{\log(6p+1)}{m+5},
\end{align*}
\fi% //////////////////////////////////////////////////////////////////////////////////////////////// \\
where $m$ satisfies~\eqref{eq:assump-on-m}. % For $p=o(\sqrt n)$, letting $m=\Theta(\sqrt{n}/p)$, we find that the  rate asymptotically satisfies
%\[R(\cC_{E})\ge \log(q-2)(1-\Theta(p/\sqrt n)).\]
 For $\log p = o(\log n_E)$, letting $m=\Theta(\log n_E)$, we find that the rate asymptotically satisfies
\[\frac{\log\|\cC_{E}\|}{n_E}\ge\log(q-2)(1-o(1)),\]
while the redundancy is at least $\Theta(n_E/\log n_E)$. We observe that a low redundancy and an asymptotic rate equal to that of $\ir_q(n_E)$ is not guaranteed for $\cC_{E}$, unlike $\codeB$, proposed in the previous section. However, $\codeB$ relies on $\cC_{E}$ to protect its syndrome as stated in Lemma~\ref{lem:last_paper}, whose proof is given in the next subsection. %We note that the rate of the code that only corrects duplications is bounded above by $\log(q-1)$.

%%%%%%%%%%%%%%%%%%%%%%%%%%%%%%%

\subsection{Proof of Lemma~\ref{lem:last_paper}}\label{subsec:proof_lemma}
To simplify the proof, instead of directly proving Lemma~\ref{lem:last_paper}, we prove the following lemma, which essentially reverses the sequences in Lemma~\ref{lem:last_paper}. Since both duplication and deduplication are symmetric operations, the lemmas are equivalent.
\begin{lemma}\label{lem:last_paper1}
Let $\bsigma = 01020$. There exists an encoder $\cE_1:\Sigma_2^{L}\to \ir_q(L')$ such that i) $\cE_1(\bu)\bsigma\in\ir_q(*)$ and ii) for any string $\bx\in \ir_q(*)$ with $\cE_1(\bu)\bsigma\bx\in \ir_q(*)$, we can recover $\bu$  from any $\bw\in R(\ddes{\leq p}(\cE_1(\bu)\bsigma\bx))$. %In other words, $\bsigma \cE_1(\bu)$ can correct  prepending any irreducible string followed by duplications and at most $p$ substitutions.
Asymptotically, $L'\le L/\log(q-2)(1+o(1))$.
\end{lemma}
\begin{IEEEproof}
Let $\bv=\cE_1(\bu)$ and $\bw\in R(\ddes{\leq p}(\bv\bsigma\bx))$. Furthermore, let $\bs$ be $|\bv|-p\cL$-prefix of $\bw$. By Lemma~\ref{lem:prefix}, we have $\bs\in\ddesded{\leq p}{\leq 2p\cL}(\bv)$. So $\bs$ can be obtained from $\bv$ through at most $3p$ $\cL$-substring edits. So if we let $\cE_1$ be an encoder for $\cC_{E}$ designed to correct $3p$ substitution errors and an infinite number of duplications, we can recover $\bu$ from $\bs$. The rate of this encoder is lower bounded by $\log(q-2)(1+o(1))$.
\end{IEEEproof}
%))))))))))))))))))))))))))))))))))))))))))))))))))))))))))))))))

\subsection{Time complexity of encoding and decoding}\label{subsec_V:C_E_time_complexity}

%\bcomment{
In this subsection, we analyze the time complexities of both the encoding and decoding algorithms for the error-correcting code in Construction~\ref{con:CMDS_Ecc_for_channels}.  Recall that we choose $T$ to be a constant and choose $\hat N=\Theta(\|\cB_{\bsigma}^m\|)$ thus satisfying $\log \hat N=\Theta(m)$% and $\hat N = \Omega((q-2)^m)$
. Also, note that %$\log n_E=\Theta(m)$
$n_E=\Theta(\hat N)$%
. Furthermore, we choose each part $\cB_{\bsigma}^m(j)$ in the partition of $\cB_{\bsigma}^m$ to be a contiguous block in the lexicographically sorted list of the elements of $\cB_{\bsigma}^m$. So the complexity of computing the mapping $\zeta_j$ is polynomial in $\|\cB_{\bsigma}^m\|$ and thus in $\hat N$.

We first discuss the complexity of the encoding. The complexity of producing the MDS codewords used in $\cC_E$ is polynomial in $\hat N$. Mapping these to sequences in $\cB_{\bsigma}^m$ is also polynomial in $\hat N$ as discussed in the previous paragraph. Hence, the encoding complexity is polynomial in $\hat N$ as well as in $n_E$.

Decoding can be performed as described in the proof of Theorem~\ref{Them:ECC_for_channels}, using the decoder described in Theorem~\ref{Theo:pIS_channel_model} and its proof. As the steps described in the proofs of these theorems are polynomial in the length of the received sequence, so is the time complexity of the decoding.

\section{Conclusion}\label{sec:conclusion}
We introduced codes for correcting any number of duplication and at most $p$ edit errors simultaneously. Recall that the set of irreducible strings is a code capable of correcting short duplication errors. %To simultaneously correct duplications and at most $p$ substitutions,
To additionally correct edit errors, we append to each irreducible sequence $\bx$ of length $n$ a vector generated through syndrome compression that enables us to distinguish confusable inputs. Given that edit and duplication errors manifest as substring edit errors, we designed a buffer and the auxiliary code in a way to enable us to recover the syndrome information from the received string. In each step of the construction, we carefully ensured that the resulting sequence is still irreducible.  %enable us to recover We carefully constructed the codewords such that the resulting sequence is itself irreducible. maintain the irreducib hen we construct an error-correcting code for duplications, at most $p$ substitutions by choosing a subset of irreducible strings of a given length. More specifically, given an irreducible input of length $n$, we start by deriving the upper bound of the number of length-$n$ confusable irreducible inputs that can generate the same output as the length-$n$ input after duplications, substitutions, and deduplications.
The additional redundancy compared to the codes correcting duplications only~\cite{jain2017duplication} is $8p(\log_q n) (1+o(1))$, with the number of edits $p$ and the alphabet size $q$ being constants, which is at most a factor of 2 away from the lowest-redundancy codes for correcting $p$ edits only~\cite{sima2020optimalCodes} and a factor of 4 away from the GV bound given in Theorem~\ref{thm:GV-bound}. %where the appended vector is used to distinguish all confusable irreducible inputs and whose length is affected by the upper bound of the number of confusable strings. Furthermore, the appended vector is itself protected against duplications and at most $p$ substitutions errors by an auxiliary error-correcting code and can be decoded correctly. Then we use it to recover the data by eliminating incorrect confusable inputs. Compared to the explicit code for duplications only~\cite{jain2017duplication}, the proposed error-correcting code corrects (duplications and) additional $\le p$ substitutions at the extra cost of roughly $8p(\log_q n) (1+o(1))$ symbols of redundancy with $q\geq 4$ and achieves the same asymptotic code rate. %Furthermore,  the proposed code outperforms the approach from~\cite{tang2021error_atmost_p} which encounters an asymptotic rate loss.
%When $p$ is a constant, both
The encoding and decoding processes have  polynomial time complexities.

The codes proposed in this work correct a wide range of errors. However, the number of edit errors is limited to be a constant. An important and interesting open problem is extending the work to correct more edits, e.g., linear in the code length. Additionally, only duplications bounded in length by three can be corrected, due to the fact that such duplications result in a regular language. So a second future direction is extending the work to correct longer duplications.
%This subsection presents that the total time complexities of both the encoding and decoding processes are polynomial, i.e., \bcomment{$O((n+p\cL)^{2p+1})$ for encoding and $O((n+p\cL)^{p+1})$ for decoding}, mainly resulting from the brute force search when $p$ is a constant with respect to $n$.

\bibliographystyle{IEEEtran}
\bibliography{references}

\begin{thebibliography}{10}
\providecommand{\url}[1]{#1}
\csname url@samestyle\endcsname
\providecommand{\newblock}{\relax}
\providecommand{\bibinfo}[2]{#2}
\providecommand{\BIBentrySTDinterwordspacing}{\spaceskip=0pt\relax}
\providecommand{\BIBentryALTinterwordstretchfactor}{4}
\providecommand{\BIBentryALTinterwordspacing}{\spaceskip=\fontdimen2\font plus
\BIBentryALTinterwordstretchfactor\fontdimen3\font minus
  \fontdimen4\font\relax}
\providecommand{\BIBforeignlanguage}[2]{{%
\expandafter\ifx\csname l@#1\endcsname\relax
\typeout{** WARNING: IEEEtran.bst: No hyphenation pattern has been}%
\typeout{** loaded for the language `#1'. Using the pattern for}%
\typeout{** the default language instead.}%
\else
\language=\csname l@#1\endcsname
\fi
#2}}
\providecommand{\BIBdecl}{\relax}
\BIBdecl

\bibitem{tang2021error_atmost_p}
Y.~Tang, H.~Lou, and F.~Farnoud, ``Error-correcting codes for short tandem
  duplications and at most $ p $ substitutions,'' in \emph{2021 IEEE
  International Symposium on Information Theory (ISIT)}.\hskip 1em plus 0.5em
  minus 0.4em\relax IEEE, 2021, pp. 1835--1840.

\bibitem{Tang2022Correcting}
Y.~Tang, S.~Wang, R.~Gabrys, and F.~Farnoud, ``Correcting multiple
  short-duplication and substitution errors,'' \emph{ISIT2022}, vol.~1, pp.
  1--6, 2022.

\bibitem{yazdi2015dna}
S.~H.~T. Yazdi, H.~M. Kiah, E.~Garcia-Ruiz, J.~Ma, H.~Zhao, and O.~Milenkovic,
  ``{{DNA}}-based storage: Trends and methods,'' \emph{IEEE Transactions on
  Molecular, Biological and Multi-Scale Communications}, vol.~1, no.~3, pp.
  230--248, 2015.

\bibitem{jain2017noise}
S.~Jain, F.~Farnoud, M.~Schwartz, and J.~Bruck, ``Noise and uncertainty in
  string-duplication systems,'' in \emph{2017 IEEE International Symposium on
  Information Theory (ISIT)}.\hskip 1em plus 0.5em minus 0.4em\relax IEEE,
  2017, pp. 3120--3124.

\bibitem{yazdi2015rewritable}
S.~H.~T. Yazdi, Y.~Yuan, J.~Ma, H.~Zhao, and O.~Milenkovic, ``A rewritable,
  random-access {DNA}-based storage system,'' \emph{Scientific reports},
  vol.~5, no.~1, pp. 1--10, 2015.

\bibitem{erlich2017dna}
Y.~Erlich and D.~Zielinski, ``{DNA} fountain enables a robust and efficient
  storage architecture,'' \emph{Science}, vol. 355, no. 6328, pp. 950--954,
  2017.

\bibitem{blawat2016forward}
M.~Blawat, K.~Gaedke, I.~Huetter, X.-M. Chen, B.~Turczyk, S.~Inverso, B.~W.
  Pruitt, and G.~M. Church, ``Forward error correction for {DNA} data
  storage,'' \emph{Procedia Computer Science}, vol.~80, pp. 1011--1022, 2016.

\bibitem{organick2018random}
L.~Organick, S.~D. Ang, Y.-J. Chen, R.~Lopez, S.~Yekhanin, K.~Makarychev, M.~Z.
  Racz, G.~Kamath, P.~Gopalan, B.~Nguyen \emph{et~al.}, ``Random access in
  large-scale {DNA} data storage,'' \emph{Nature biotechnology}, vol.~36,
  no.~3, pp. 242--248, 2018.

\bibitem{lee2019terminator}
H.~H. Lee, R.~Kalhor, N.~Goela, J.~Bolot, and G.~M. Church, ``Terminator-free
  template-independent enzymatic {{DNA}} synthesis for digital information
  storage,'' \emph{Nature communications}, vol.~10, no.~1, pp. 1--12, 2019.

\bibitem{jain2017duplication}
S.~Jain, F.~Farnoud, M.~Schwartz, and J.~Bruck, ``Duplication-correcting codes
  for data storage in the {{DNA}} of living organisms,'' \emph{IEEE
  Transactions on Information Theory}, vol.~63, no.~8, pp. 4996--5010, 2017.

\bibitem{shipman2017}
S.~L. Shipman, J.~Nivala, J.~D. Macklis, and G.~M. Church,
  ``\BIBforeignlanguage{en}{{{CRISPR}}\textendash{{Cas}} encoding of a digital
  movie into the genomes of a population of living bacteria},''
  \emph{\BIBforeignlanguage{en}{Nature}}, vol. 547, no. 7663, pp. 345--349,
  Jul. 2017.

\bibitem{kovavcevic2018asymptotically}
M.~Kova{\v{c}}evi{\'c} and V.~Y. Tan, ``Asymptotically optimal codes correcting
  fixed-length duplication errors in {{DNA}} storage systems,'' \emph{IEEE
  Communications Letters}, vol.~22, no.~11, pp. 2194--2197, 2018.

\bibitem{yehezkeally2019reconstruction}
Y.~Yehezkeally and M.~Schwartz, ``Reconstruction codes for {DNA} sequences with
  uniform tandem-duplication errors,'' \emph{IEEE Transactions on Information
  Theory}, vol.~66, no.~5, pp. 2658--2668, 2020.

\bibitem{tang2020single}
Y.~{Tang}, Y.~{Yehezkeally}, M.~{Schwartz}, and F.~{Farnoud}, ``Single-error
  detection and correction for duplication and substitution channels,''
  \emph{IEEE Transactions on Information Theory}, vol.~66, no.~11, pp.
  6908--6919, 2020.

\bibitem{lenz2019coding}
A.~{Lenz}, P.~H. {Siegel}, A.~{Wachter-Zeh}, and E.~{Yaakobi}, ``Coding over
  sets for {{DNA}} storage,'' \emph{IEEE Transactions on Information Theory},
  vol.~66, no.~4, pp. 2331--2351, 2020.

\bibitem{cai2019optimal}
K.~Cai, Y.~M. Chee, R.~Gabrys, H.~M. Kiah, and T.~T. Nguyen, ``Optimal codes
  correcting a single indel/edit for {{DNA}}-based data storage,'' \emph{arXiv
  preprint arXiv:1910.06501}, 2019.

\bibitem{elishco2019bounds}
O.~Elishco, R.~Gabrys, and E.~Yaakobi, ``Bounds and constructions of codes over
  symbol-pair read channels,'' \emph{IEEE Transactions on Information Theory},
  vol.~66, no.~3, pp. 1385--1395, 2020.

\bibitem{lenz2020coding}
A.~Lenz, Y.~Liu, C.~Rashtchian, P.~H. Siegel, A.~Wachter-Zeh, and E.~Yaakobi,
  ``Coding for efficient {{DNA}} synthesis,'' in \emph{IEEE International
  Symposium on Information Theory (ISIT)}.\hskip 1em plus 0.5em minus
  0.4em\relax IEEE, 2020, pp. 2885--2890.

\bibitem{gabrys2020mass}
R.~{Gabrys}, S.~{Pattabiraman}, and O.~{Milenkovic}, ``Mass error-correction
  codes for polymer-based data storage,'' in \emph{IEEE International Symposium
  on Information Theory (ISIT)}, 2020, pp. 25--30.

\bibitem{jain2020coding}
S.~{Jain}, F.~{Farnoud}, M.~{Schwartz}, and J.~{Bruck}, ``Coding for optimized
  writing rate in {{DNA}} storage,'' in \emph{IEEE International Symposium on
  Information Theory (ISIT)}, 2020, pp. 711--716.

\bibitem{kiah2020coding}
H.~M. {Kiah}, T.~{Thanh Nguyen}, and E.~{Yaakobi}, ``Coding for sequence
  reconstruction for single edits,'' in \emph{IEEE International Symposium on
  Information Theory (ISIT)}, 2020, pp. 676--681.

\bibitem{yehezkeally2020uncertainty}
Y.~{Yehezkeally} and M.~{Schwartz}, ``Uncertainty of reconstructing multiple
  messages from uniform-tandem-duplication noise,'' in \emph{IEEE International
  Symposium on Information Theory (ISIT)}, 2020, pp. 126--131.

\bibitem{nguyen2020constrained}
T.~T. Nguyen, K.~Cai, K.~A.~S. Immink, and H.~M. Kiah, ``Constrained coding
  with error control for {{DNA}}-based data storage,'' in \emph{IEEE
  International Symposium on Information Theory (ISIT)}.\hskip 1em plus 0.5em
  minus 0.4em\relax IEEE, 2020, pp. 694--699.

\bibitem{sima2020robust}
J.~Sima, N.~Raviv, and J.~Bruck, ``Robust indexing-optimal codes for {{DNA}}
  storage,'' in \emph{IEEE International Symposium on Information Theory
  (ISIT)}.\hskip 1em plus 0.5em minus 0.4em\relax IEEE, 2020, pp. 717--722.

\bibitem{chee2020efficient}
Y.~M. Chee, J.~Chrisnata, H.~M. Kiah, and T.~T. Nguyen, ``Efficient
  encoding/decoding of {GC}-balanced codes correcting tandem duplications,''
  \emph{IEEE Transactions on Information Theory}, vol.~66, no.~8, pp.
  4892--4903, 2020.

\bibitem{Tang2021CorrectingDeletion}
Y.~Tang and F.~Farnoud, ``Correcting deletion errors in {DNA} data storage with
  enzymatic synthesis,'' in \emph{2021 IEEE Information Theory Workshop (ITW)},
  2021, pp. 1--6.

\bibitem{Tang2021Noisy}
------, ``Error-correcting codes for noisy duplication channels,'' \emph{IEEE
  Transactions on Information Theory}, vol.~67, no.~6, pp. 3452--3463, 2021.

\bibitem{jain2017capacity}
S.~Jain, F.~Farnoud, and J.~Bruck, ``Capacity and expressiveness of genomic
  tandem duplication,'' \emph{IEEE Transactions on Information Theory},
  vol.~63, no.~10, pp. 6129--6138, 2017.

\bibitem{kovavcevic2019codes}
M.~Kova{\v{c}}evi{\'c}, ``Codes correcting all patterns of tandem-duplication
  errors of maximum length 3,'' \emph{arXiv preprint arXiv:1911.06561}, 2019.

\bibitem{chee2019deciding}
Y.~M. Chee, J.~Chrisnata, H.~M. Kiah, and T.~T. Nguyen, ``Deciding the
  confusability of words under tandem repeats in linear time,'' \emph{ACM
  Transactions on Algorithms (TALG)}, vol.~15, no.~3, pp. 1--22, 2019.

\bibitem{tang2020error}
Y.~Tang and F.~Farnoud, ``Error-correcting codes for short tandem duplication
  and substitution errors,'' in \emph{IEEE International Symposium on
  Information Theory (ISIT)}.\hskip 1em plus 0.5em minus 0.4em\relax IEEE,
  2020, pp. 734--739.

\bibitem{tang2019error}
------, ``Error-correcting codes for noisy duplication channels,'' in
  \emph{2019 57th Annual Allerton Conference on Communication, Control, and
  Computing (Allerton)}.\hskip 1em plus 0.5em minus 0.4em\relax IEEE, 2019, pp.
  140--146.

\bibitem{Tang2021ECC_Edit}
------, ``Error-correcting codes for short tandem duplication and edit
  errors,'' \emph{IEEE Transactions on Information Theory}, vol.~68, no.~2, pp.
  871--880, 2022.

\bibitem{marcusintroduction2001}
B.~H. Marcus, R.~M. Roth, and P.~H. Siegel, ``An introduction to coding for
  constrained systems,'' \emph{Lecture notes}, 2001.

\bibitem{sima2020syndrome}
J.~Sima, R.~Gabrys, and J.~Bruck, ``Syndrome compression for optimal redundancy
  codes,'' in \emph{2020 IEEE International Symposium on Information Theory
  (ISIT)}.\hskip 1em plus 0.5em minus 0.4em\relax IEEE, 2020, pp. 751--756.

\bibitem{sima2020optimal}
J.~Sima and J.~Bruck, ``On optimal $k$-deletion correcting codes,'' \emph{IEEE
  Transactions on Information Theory}, vol.~67, no.~6, pp. 3360--3375, 2020.

\bibitem{sima2020optimalCodes}
J.~Sima, R.~Gabrys, and J.~Bruck, ``Optimal codes for the q-ary deletion
  channel,'' in \emph{2020 IEEE International Symposium on Information Theory
  (ISIT)}.\hskip 1em plus 0.5em minus 0.4em\relax IEEE, 2020, pp. 740--745.

\bibitem{sima2020optimal_Systematic}
------, ``Optimal systematic $t$-deletion correcting codes,'' in \emph{2020
  IEEE International Symposium on Information Theory (ISIT)}.\hskip 1em plus
  0.5em minus 0.4em\relax IEEE, 2020, pp. 769--774.

\end{thebibliography}

%%%%%%%%%%%%%%%%%%%%%%%%%%%%%%%%%%%%%%%%%%%%%%%%%%%%%%%%%%%%%%%%%%%%%%%%%%%%%%%%%%%%%%%%%%%%%%%%%%
%%%%%%%%%%%%%%%%%%%%%%%%%%%%%%%%%%%%%%%%%%%%%%%%%%%%%%%%%%%%%%%%%%%%%%%%%%%%%%%%%%%%%%%%%%%%%%%%%

\begin{appendices}

\section{Proof of Lemma~\ref{lem:dominance}}\label{app:lem_dom}
\dominance*
To prove Lemma~\ref{lem:dominance}, we start with the definition of dominance between two sequences from \cite{Tang2021ECC_Edit}.

\begin{definition}\label{def:dominance}
Let $\bs$ and $\bar\bs$ be strings of length $n$, and let $A$ be the set of symbols in $\bs$ and $\bar A$ the set of symbols in $\bar\bs$. We say that $\bs$ \textit{dominates} $\bar\bs$ if there exists a function $\eta:A\to \bar A$ such that $\bar\bs=\eta(\bs)$, where $\eta(\bs) = \eta(s_1)\dotsm \eta(s_n)$. Furthermore, a set $U$ of strings dominates a set $T$ if there is a single mapping $\eta$ such that for each string $\bt\in T$ there is a string $\bu\in U$ such that $\bt=\eta(\bu)$.
\end{definition}

 For example, $0102$ dominates $1212$ (using the mapping $\eta(0)=1,\eta(1)=2,\eta(2)=2$) but $0102$ does not dominate $0010$. The string $012\dotsm k$ dominates any string of length $k+1$.

 We recall an auxiliary lemma showing properties of dominance from~\cite{Tang2021ECC_Edit}, along with two other auxiliary lemmas that are used to simplify the proof of Lemma~\ref{lem:dominance}.

% \rcomment{If this appears in a previous paper (excluding the 2 conference papers that are part of this paper) there should be a citation here.}
% \bcomment{Yuanyuan: I fixed it.}

 \begin{lemma}\label{lemma:domonance_property_dedup}
 (\cite[Lemma 1]{Tang2021ECC_Edit})
 Assume there are two strings $\bs, \bar \bs$ with $\bs$ dominating $\bar \bs$. \begin{enumerate}
     \item    Suppose we apply the same duplication in both $\bs$ and $\bar\bs$ (that is, in the same position and with the same length). Let the resulting strings be $\bs'$ and $\bar\bs'$, respectively. Then $\bs'$ dominates $\bar\bs'$.
     \item  If a deduplication is possible in $\bs$, a deduplication in the same position and with the same length is possible in $\bar\bs$. Let the result of applying this deduplication to $\bs$ and $\bar \bs$ be denoted by $\bs'$ and $\bar\bs'$, respectively. Then $\bs'$ dominates $\bar\bs'$.
 \end{enumerate}
 \end{lemma}

% ////////////////////////////////////////////////////////////////////////////////////////////////

%%%%%%%%%%%%%%%%%%%%%%%%%%%%%
%%%% Comment out below
%%%%%%%%%%%%%%%%%%%%%%%%%%%%%%%%%
%\iffalse

%\rcomment{To do: Indicate that inserted/substituted symbol belong to the same alphabet.} \bcomment{ I highlight that after \eqref{eq:decomposition}.}

\begin{lemma}\label{lem:aux_dominante}
%bcomment{
Let $\bar \bs$ be a string over $\bar{\Sigma}$ and $\bs$ a string over $\Sigma$ such that $\bs$ dominates $\bar\bs$.  Let the number of distinct symbols in $\bar\bs$ and $\bs$ be denoted $\bar q_{s}$ and $q_{s}$, respectively, and suppose $\|\Sigma\| \geq \|\bar\Sigma\|+(q_{s}-\bar q_s)$. Then $\ddes{p}( \bs) \subseteq \Sigma^{*}$ dominates $\ddes{p}(\bar\bs) \subseteq \bar \Sigma^*$. In other words, there is a mapping $\eta:\Sigma \to \bar \Sigma$ that for any $\bar \by \in \ddes{p}(\bar\bs)\subseteq \bar \Sigma^*$, there exists $\by \in \ddes{p}(\bs)\subseteq \Sigma^{*}$ such that $\bar\by = \eta(\by)$.
\end{lemma}

Before proving the lemma, we provide an example with multiple short duplications and a substitution error, where the duplicated substrings are marked with underlines and the substituted symbols are in red.

Let $\Sigma=\{0,1,2,3,4\}$ and $\bar\Sigma=\{0,1,2,3\}$. Suppose $\bs=012$ and $\bar \bs=010$ with $q_s=3$ and $\bar q_s=2$. The mapping $\eta(0)=0$, $\eta(1)=1$, and $\eta(2)=0$, shows that  $\bs$ dominates $\bar \bs$, i.e., $\bs=012\to \bar\bs=010$.

    Let $\bar \by_1=010 \underline{010} \underline{ 010}\in \ddeo( \bar \bs)$. Then there exists $\by_1=012 \underline{012 012}\in \ddeo(  \bs)$ dominating $\bar \by_1$, via the same mapping~$\eta$.

    Next, assume $\bar \by_2=010 01\rcomment{2} 010$ is generated from $\bar \by_1$  by a substitution $0 \to \rcomment{2}$. Then $\by_2=012 01\rcomment{3} 012$, obtained from $\by_1$ after a substitution $2\to \rcomment{3}$ in the same position, dominates $\bar \by_2$, via the mapping $\eta$ extended by  $\eta(3)=2$. %Note that, if  $|\Sigma|-q_{s}\geq |\bar\Sigma|-\bar q_s$, we can always find a symbol $b\in \Sigma \setminus \{[0,q_s-1]\}$ for each new substituted symbol $a\in \bar \Sigma\setminus \{[0,\bar q_s-1]\}$ such that $a=\eta(b)$.

%\iffalse
\begin{IEEEproof}[Proof of Lemma~\ref{lem:aux_dominante}]
 %\bcomment{
 Without loss of
 generality, assume that $\bar\Sigma = \{0,1,\dotsc,\|\bar\Sigma\|-1\}$ and that the symbols appearing in $\bar\bs$ are $0,1,\dotsc,\bar q_s-1$, where $\bar q_s\le\|\bar\Sigma\|$. Similar statements hold for $\Sigma,\bs,q_s$. By assumption, there exists some mapping $\eta:\{0,\dotsc, q_s-1\}\to \{0,\dotsc,\bar q_s-1\}$ showing that $\bs$ dominates $\bar\bs$. %Suppose $p\leq \vert\Sigma \vert-q_s$, we extend this mapping by assigning $\eta i \to i-(q_s-\bar q_s)$ for $q_s\le i< |\Sigma|$ to construct $\eta:\Sigma\to\bar\Sigma$.
Since $\Vert\Sigma \Vert-q_s\geq \Vert\bar \Sigma \Vert-\bar q_s$, we may extend $\eta$ by mapping symbols in $\Sigma$  not occurring in $\bs$ to symbols in $\bar\Sigma$ not occurring in $\bar s$. Specifically, we assign $\eta(i)=i-(q_s-\bar q_s)\in \bar \Sigma$ for $i\in \{q_s, q_s+1,\dotsc,\|\Sigma\|-1\}\subseteq \Sigma$ to construct $\eta:\Sigma\to\bar\Sigma$
.

Let the sequence of errors transforming $\bar\bs$ to $\bar\by$ be denoted by $\bar T_j, j = 1,\dotsc,k$ and let $\bar \by_j = \bar T_j(\bar \by_{j-1})$ with $\bar \by_0 = \bar \bs$ and $\bar\by = \bar\by_k$. We will find a corresponding sequence $(T_j)$, where each $T_j$ has the same type of error as  $\bar T_j$, and define $\by_j=T_j(\by_{j-1})$. %Note that the number of substitutions satisfies $p\leq \vert \Sigma \vert-q_s$.
We prove that for each $j$, we have $\bar\by_j = \eta(\by_j)$. The claim holds for $j=0$ by assumption. Suppose it holds for $j-1$. We show that it also holds for $j$. % with $j\leq \vert \Sigma \vert-q_s$.
If $\bar T_j$ is a duplication, by Lemma~\ref{lemma:domonance_property_dedup}.1), then we choose $T_j$ to be a duplication of the same length in the same position. %\deleted{If $\bar T_j$ is a deletion, then $T_j$ is also a deletion. If $\bar T_j$ inserts a symbol $a\in\bar\Sigma$ into $\bar\by_{j-1}$, we choose $T_j$ to be an insertion of some symbol $b$ into $\by_{j-1}$ such that $f(b)=a$.}
If $\bar T_j$ substitutes some symbol in $\bar\by_{j-1}$ with $a\in \bar\Sigma$, then $T_j$ substitutes the symbol in the same position in $\by_{j-1}$ with a symbol $b\in \Sigma$ such that $\eta(b)=a$.  It then follows that $\bar\by_j = \eta(\by_j)$ for each $\bar \by_j$. Therefore, we have $\ddes{p}( \bs) \subseteq \Sigma^{*}$ dominates $\ddes{p}(\bar\bs) \subseteq \bar \Sigma^*$.
%}
\end{IEEEproof}
%\fi

\begin{lemma}\label{lem:roots_bound_dominance}
If a set of strings $Y$ dominates a second set $\bar Y$, then $||R(\bar Y)||\le ||R(Y)||$.
\end{lemma}

%The lemma is proved in Appendix~\ref{appendix:lem:roots_bound_dominance}.

%\iffalse

%Based on the two properties ahead, we present the proof.

\begin{IEEEproof}
Suppose $Y$ dominates $\bar Y$ via a mapping $\eta:\Sigma \to \bar \Sigma$. Then, for each $\bar \by\in \bar Y$, there exists some $\by \in Y$ such that $\bar \by=\eta(\by)$. For $\bar \by\in \bar Y$, define $\eta^{-1}(\bar\by)$ as the lexicographically-smallest sequence among $\{\by\in Y: \eta(\by)=\bar\by\}$. Furthermore, define $Y'=\{\eta^{-1}(\bar\by):\bar\by\in\bar Y\}$ and note that $Y'\subseteq Y$. With this definition, $Y'$ dominates $\bar Y$ and $\eta$ is a bijection between the two sets. We have $\Vert \bar Y\Vert =  \Vert Y'\Vert \leq \Vert Y\Vert $. Also, as $Y'\subseteq Y$, we have $\|R(Y')\|\le \|R(Y)\|$.

To prove the lemma, we show that $\|R(\bar Y)\|\le \|R(Y')\|$. It suffices to prove that if $\bar \by_1, \bar \by_2 \in \bar Y$ have distinct roots, then $\by_1,\by_2\in Y'$, where $\by_1=\eta^{-1}(\bar\by_1)$ and $\by_2=\eta^{-1}(\bar\by_2)$, also have distinct roots.

Suppose, on the contrary, that $\by_1,\by_2$ do not have distinct roots, i.e., $R(\by_1)=R(\by_2)$. Let $T_1$ and $T_2$ represent the sequences of deduplications on $\by_1$ and $\by_2$ that produce their roots, i.e., $R(\by_1)=T_1(\by_1)$ and $R(\by_2)=T_2(\by_2)$. Based on the Lemma~\ref{lemma:domonance_property_dedup}.2) above, there exist two corresponding sequences of deduplications $\bar T_1$ and $\bar T_2$ %over $\bar \by_1$ and $\bar \by_2$
such that $\bar T_1(\bar \by_1)=\eta(R(\by_1))$ and $\bar T_2(\bar \by_2)=\eta(R(\by_2))$. If $R(\by_1)=R(\by_2)$, %$\bar y_2=\eta(\by_2)$
%based on the specific mapping $\eta$, we have
then $\bar T_1(\bar \by_1)=\bar T_2(\bar \by_2)$. But by the uniqueness of the root, $R(\bar\by_1)=R(\bar T_1(\bar \by_1))$ and $R(\bar\by_2)=R(\bar T_2(\bar \by_2))$. So $R(\bar \by_1)=R(\bar \by_2)$. But this contradicts the assumption. Hence, the roots of $\by_1$ and $\by_2$ are distinct.
%))))))))))))))))))))))))))))))))
\end{IEEEproof}
%\fi

With Lemma~\ref{lem:aux_dominante} and Lemma~\ref{lem:roots_bound_dominance} in hand, we prove Lemma~\ref{lem:dominance} in the following.
\begin{IEEEproof}[Proof of Lemma~\ref{lem:dominance}]
Let $\bs=01234$. If $\bt$ is the empty string, the claim is trivial. So in the rest of the proof, we assume $\bt$ is not empty. Based on Definition~\ref{def:dominance}, $\bs$ dominates $\bt$ for any $\bt\in \Sigma_q^{5}\setminus\{\Lambda\}$. Let $q_t$ denote the number of distinct symbols in $\bt$ and note that there are 5 distinct symbols in $\bs$. By Lemma~\ref{lem:aux_dominante}, with $p=1$, $\ddes{1}(\bs)\subseteq \Sigma_{q+4}^*$ dominates $\ddes{1}(\bt)\subseteq \Sigma_q^*$ for any $\bt\in \Sigma_q^5$ since $q+4\ge q + (5-q_t)$ as $q_t\ge 1$. % Based on Lemma~\ref{lem:roots_bound_dominance}, we have $\Vert R(\ddes{1}(\bs))\Vert \geq \Vert R(\ddes{1}(\bt))\Vert$ for each $\bt\in \Sigma_q^5$.
%Therefore, we prove Lemma~\ref{lem:dominance}.
Applying Lemma~\ref{lem:roots_bound_dominance} to $\ddes{1}(\bs)$ and $\ddes{1}(\bt)$ completes the proof.
\end{IEEEproof}
%\rcomment{Yuanyuan: the proof seems incomplete. We may also need to mention  at the end of the proof. }

%\fi

%%%%%%%%%%%%%%%%%%%%%%%%%%%%%%%%%%%%%%%%%%%

\section{Proof of Lemma~\ref{lem:symmetry}}\label{app:symmetry}
\symmetry*
%\iffalse
\begin{IEEEproof}
Define $h(a)=4-a$ for $a\in \Sigma_5$ and $h(\bu) =  h(u_n) h(u_{n-1}) \dotsm h(u_{1})$ for $\bu\in\Sigma_5^n$. Furthermore, for $S\subseteq \Sigma_5^*$, define $h(S)=\{h(\bu):\bu\in S\}$. Note that $h$ is its own inverse. We claim that $h$ has the following properties, to be proved later:
\begin{enumerate} %[label=\alph*]
    \item \label{itm:p1} For $\bs,\bt\in\Sigma_5^*$, $\bs$ is a prefix of $\bt$ if and only if $h(\bs)$ is a suffix of $h(\bt)$.
    \item \label{itm:p2} For $\bt\in\Sigma_5^*$, $\ddeo(h(\bt))=h(\ddeo(\bt))$.
    \item \label{itm:p3} For $S\subseteq\Sigma_5^*$, $R(h(S))=h(R(S))$.
\end{enumerate}

By definition, if $\bu\in U$ then $\bu$  is a prefix  of  some $\bx\in \ddeo(01234)$.  Then, by Property~\ref{itm:p1}, $h(\bu)$ is a suffix of $h(\bx)$. By setting $\bt=01234$, it follows from Property~\ref{itm:p2} that $\ddeo(01234)= h(\ddeo(01234))$, and thus $h(\bx)\in \ddeo(01234)$. Hence, $h(\bu)$ is in $V$. Similarly, we can show that if $\bv\in V$, then $h(\bv)\in U$. As $h$ is its own inverse, we have $V=h(U)$ and $\|U\|=\|V\|$. Applying  Property~\ref{itm:p3} with $S=U$ yields $R(V)=h(R(U))$ and $\|R(V)\|=\|R(U)\|$.

We now prove Properties~\ref{itm:p1}-\ref{itm:p3}. Property~\ref{itm:p1} follows from the definition of $h$. Property~\ref{itm:p2} follows from the observation that if $\bx'$ is obtained from $\bx$ via a duplication, then $h(\bx')$ can be obtained from $h(\bx)$ via a duplication, i.e., the relationship represented by $h$ is maintained under duplication. To prove Property~\ref{itm:p3}, it suffices to show that $R(h(\bt))=h(R(\bt))$ for $\bt\in\Sigma_5^*$, which holds as $h$ is maintained under deduplication.

%>>>>>>>>>>>>>>>>>>>>>>>>>>>>>>>>>>>>>>>>>>>>>>>
\end{IEEEproof}
%\fi

%\rcomment{Yuanyuan: the proof of three properties seem to be very short.}

%%%%%%%%%%%%%%%%%%%%%%%%%%%%%%%%%%%%%%%%%%%%%%%%%%%%
\section{Proof of Lemma~\ref{lem:prefix}}\label{app:prefix}
\prefix*

\begin{IEEEproof}
Based on  Theorem~\ref{Theo:root_L_17}, $\bw$ can be considered as being generated from $\bx\br$ by at most $p$ $\cL$-substring edits. Let $j$ be the last symbol of $\bx$ not affected by a substring edit (i.e., it is not deleted by a substring edit, but it may be shifted). Suppose $t\le p$ substring edits occur before $x_j$ and at most $p-t$ after $x_j$. Then, $j\in[n-(p-t)\cL,n]$. The symbol $x_j$ appears as the $i$th symbol of $\bw$ for some $i\in[j-t\cL,j+t\cL]$ %\bcomment{? $i\in[j-t\cL,j+t\cL]$ or $i\in[n-p\cL,n+t\cL]$}
. Then, $\bw_{[i]}\in R(\ddes{t}(\bx_{[j]}))$. It follows that $\bv\in R(\ddes{t}(\bx))$ for $\bv = \bw_{[i]}\bx_{[j+1,n]}$. As $i\ge j-t\cL$ and $j\ge n-(p-t)\cL$, we have $n-p\cL\le i$. Hence, $\bs = \bw_{[n-p\cL]}$ is a prefix of $\bw_{[i]}$ and thus also a prefix of $\bv$. Specifically, $\bs$ can be obtained from $\bv$ by a suffix deletion of length
\begin{align*}
    |\bv|-(n-p\cL) &= i+(n-j)-(n-p\cL) \\
    &\le n+t\cL + (p-t)\cL-(n-p\cL)\\
    &=2p\cL.
\end{align*} As $\bv\in\ddes{\le p}(\bx)$, we have $\bs\in \ddesded{\leq p}{\leq 2p\cL}(\bx)$.
\end{IEEEproof}

%]]]]]]]]]]]]]]]]]]]]]]]]]]]]]]]]]]]]]]]]]

%%%%%%%%%%%%%%%%%%%%%%%%%%%%%%%%%%%%%%%%%%%%%%%%%%%%%%%%%%
\section{Proof of Lemma~\ref{lem:buffer}}\label{Appdix:lemma:length_buffer}
\buffer*

Before proving Lemma~\ref{lem:buffer}, we recall from~\cite{jain2017duplication} that $\ir_q(*)$ is a regular language whose graph $G_q=(V_q, \xi_q)$ is a subgraph of the De Bruijn graph. The vertex set $V_q$ consists of 5-tuples $a_1a_2a_{3}a_{4}a_5 \in \ir_q(5)$ that do not have any repeats (of length at most $4$). There is an edge from $a_{1}a_{2}a_{3}a_{4}a_{5} \to a_2a_3a_4a_5a_6$ if $a_{1}a_{2}a_{3}a_{4}a_{5}a_{6}$ belongs to $\ir_q(6)$. The label for this edge is $a_6$. The label for a path is the 5-tuple representing its starting vertex concatenated with the labels of the subsequent edges. %\bcomment{Furthermore, the length of a path is defined as the number of edges in the path.} %Furthermore, each irreducible string is represented by a unique path. The length of a path is defined as the number of edges in the path. %\rcomment{[Yuanyuan: is this claim requires a proof?]}
% In graph theory, the length of a path is by definition the number of edges in the path.
The proof below is similar to that of~\cite[Theorem~15]{Tang2021ECC_Edit} and is presented here for completeness.
%\rcomment{Yuanyuan: why do you replace $a_1$ with $a'$ above?}
%\rcomment{This proof is very similar to \cite[Theorem~15]{Tang2021ECC_Edit}. What is the purpose of including it? What is the difference?}
%\rcomment{Yuanyuan: The proof here is to show that there exists some integer $c_q$ such any vertex can reach $\bsigma=01020$ by a path over the subgraph with length upper bounded by $c_q+5$, where $5$ represents the last $5$ edges reaching $\bsigma$, and $c_q$ represents the maxi-min value of the length of $\bb_{\bx}$ for a given $q$. since the label of a path contains the first vertex. Compared to \cite[Theorem~15]{Tang2021ECC_Edit}, the main difference is that the paths from other vertices to $\bsigma$ do not need to avoid the vertex $\bsigma$. And the integer $c_q$ is defined to represent the maxi-min of length $\bb_\bx$.}
\begin{IEEEproof}
Given $\bx\in \ir_q(n)$ and $q\geq 3$, $\bx$ can be represented by a path over the graph $G_q$, ending at the vertex $\bx_{[n-4:n]}$. Furthermore, $\bsigma=01020$ can be considered as a vertex in $G_q$ since $\bsigma\in \ir_q(5)$. Let us assume for the moment that $q\ge 6$. Based on~\cite[Lemma~14]{Tang2021ECC_Edit}, each vertex has at least $q-2$ outgoing edges. So from each vertex, there is at  least one outgoing edge whose label is equal to either $3,4,$ or $5$. So, starting from $\bx_{[n-4:n]}$, we may arrive at some vertex with label $\bb_\bx\in\{3,4,5\}^5$ in 5 steps. Furthermore,  $\bb_\bx\bsigma$ is irreducible as both $\bb_\bx$ and $\bsigma$ are irreducible and have no symbols in common. Hence, there is a path of length 5 from $\bb_\bx$ to $\bsigma$ in $G_q$. So there is a path in $G_q$ with label $\bx\bb_\bx\bsigma$, implying that $\bx\bb_\bx\bsigma$ is irreducible. We further have $c_q=|\bb_\bx|=5$. For $q\in\{3,4,5\}$, we have verified computationally that, for any choice of $\bx_{[n-4:n]}$, there exists a path from  $\bx_{[n-4:n]}$ to  $\bsigma$ of length $c_q+5$, with the value of $c_q$ as given in the lemma. Denoting the label of this path as $\bb_\bx\bsigma$ gives us the sequence $\bb_\bx$ of length $c_q$, with $\bx\bb_\bx\bsigma$ being irreducible.
%))))))))))))))))))))))))))
\end{IEEEproof}

\begin{figure}[h]
    \centering
    \includegraphics[width=0.35\textwidth]{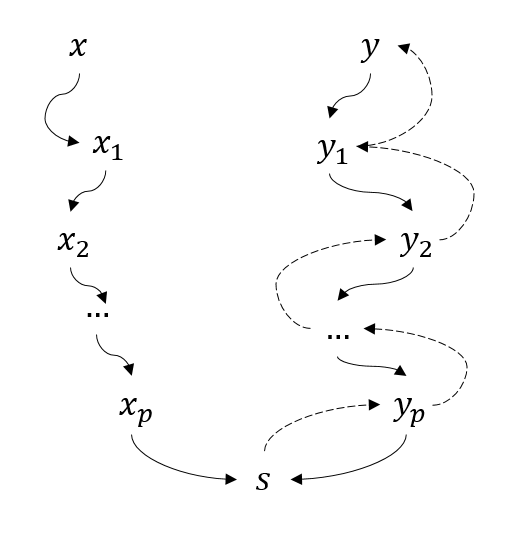}
    \caption{$\bs$ results from passing $\bx$ and $\by$ through a concatenation of $p$ DSD(1) channels and a channel deleting a suffix of length at most $2p\cL$ (c.f.\ Figure~\ref{fig:dup_subs_ded_channel_and_equivalent_channel}).% equivalent channel by concatenating $p$  sub-channels and one channel with at most one suffix-burst deletion, where each sub-channel consists of duplications, at most one substitution, and deduplications. Note that given $\bx\in \ir_q(n)$ and $\by \in  B_{\ir}^{\leq p,\leq 2 p\cL}(\bx)$, let $\bs\in \ddesded{\leq p}{\leq 2p\cL}(\bx)\cap \ddesded{\leq p}{\leq 2p\cL}(\by)$. Furthermore, $\bx_i$ (resp.\ $\by_i$) is over sequences that can result from $\bx$ (resp.\ $\by$) passing through the concatenation of $i$ DSD(1) channels for $i\in[p]$. Note that $\bx_0=\bx$ and $\by_0=\by$. Then $\bs$ is a descendant generated from both $\bx_p$ and $\by_p$ by deleting a suffix of length at most $2p\cL$.
    }
    \label{fig:equivalent_channel_tail_burst_deletion}
\end{figure}

\section{Example for Theorem~\ref{Theo:root_L_17}}\label{app:example}
\rootpL*

The following example illustrates the theorem.

\begin{example}\label{Example:L_substring_edit_errors}
Let the alphabet be
$\Sigma_4=\{0,1,2,3\}$ and $p=2$. We take the input $\bx$ to be irreducible, i.e., $R(\bx)=\bx$. By passing through the channel, $\bx$ suffers multiple duplications and $2$ symbol substitutions, resulting in  $\by\in \ddes{2}(\bx)$. We show the difference between $R(\bx)$ and $R(\by)$ for two possible input-output pairs. Below, substrings added via duplication  are marked with underlines, while substituted symbols are red and bold.

First, we provide an example where $R(\by)$ can be obtained from $R(\bx)$ via non-overlapping substring edits:
    \begin{align*}
        \bx&=321 031 3230121321,\\
        \by&=321 \underline{32\rcomment{\mathbf0}321} 031\underline{31} 32\underline{132}3\rcomment{\mathbf2}12132\underline{132}1,\\
        R(\bx)&=\underbrace{321}_{\balpha_0}\underbrace{\phantom{\Lambda}}_{\bbeta_1}\underbrace{031}_{\balpha_1} \underbrace{3230121}_{\bbeta_2}\underbrace{321}_{\balpha_2},\\
        R(\by)&=\underbrace{321}_{\balpha_0}\underbrace{32\rcomment{\mathbf0}321}_{\bbeta'_1}\underbrace{031}_{\balpha_1} \underbrace{\phantom{\Lambda}}_{\bbeta'_2}\underbrace{321}_{\balpha_2},
    \end{align*}
    where the errors are $\bbeta_1=\Lambda\to \bbeta'_1$ and $\bbeta_2\to \bbeta'_2=\Lambda$.

In the second case, the two edits overlap, leading to a single substring substitution:%\rcomment{F: I'm not sure the underlines below are correct.}\bcomment{Yuanyuan:I missed a underline. the first substitution is to substitute a duplication symbol $3$.}
    \begin{align*}
        \bx&=132 031 230,\\
        \by&=132 \underline{32} 03\underline{\rcomment{\mathbf2}}1 \underline{32 \rcomment{\mathbf0} 321}230\underline{23 023 0},\\
        R(\bx)&=\underbrace{13203}_{\balpha_0}\underbrace{\phantom{\Lambda}}_{\bbeta} %\underbrace{3230121}_{\bbeta_2}
               \underbrace{1 230}_{\balpha_1}\\
        R(\by)&=\underbrace{13203}_{\balpha_0}\underbrace{\rcomment{\mathbf2}1 32 \rcomment{\mathbf0} 32}_{\bbeta'}\underbrace{1 230}_{\balpha_1}.
    \end{align*}
\end{example}

%%%%%%%%%%%%%%%%%%%%%%%%%%%%%%%%%%%%%%%%%%%%%%%%%%%%%%%%%%%

\section{Proof of Lemma~\ref{lem:prefix_B}}\label{sec:prefix_B}
%\prefix*
\prefixB*

\iffalse
\begin{figure}[h]
    \centering
    \includegraphics[width=0.35\textwidth]{Figs_Syn/p_subs_burst_del_p_channel.PNG}
    \caption{$\bs$ results from passing $\bx$ and $\by$ through a concatenation of $p$ DSD(1) channels and a channel deleting a suffix of length at most $2p\cL$ (c.f.\ Figure~\ref{fig:dup_subs_ded_channel_and_equivalent_channel}).% equivalent channel by concatenating $p$  sub-channels and one channel with at most one suffix-burst deletion, where each sub-channel consists of duplications, at most one substitution, and deduplications. Note that given $\bx\in \ir_q(n)$ and $\by \in  B_{\ir}^{\leq p,\leq 2 p\cL}(\bx)$, let $\bs\in \ddesded{\leq p}{\leq 2p\cL}(\bx)\cap \ddesded{\leq p}{\leq 2p\cL}(\by)$. Furthermore, $\bx_i$ (resp.\ $\by_i$) is over sequences that can result from $\bx$ (resp.\ $\by$) passing through the concatenation of $i$ DSD(1) channels for $i\in[p]$. Note that $\bx_0=\bx$ and $\by_0=\by$. Then $\bs$ is a descendant generated from both $\bx_p$ and $\by_p$ by deleting a suffix of length at most $2p\cL$.
    }
    \label{fig:equivalent_channel_tail_burst_deletion}
\end{figure}
\fi

\begin{IEEEproof}
    The proof is similar to that of Theorem~\ref{Theo:max_root_set_p_subs}, but also takes into account the effect of the suffix deletions, as shown in Figure~\ref{fig:equivalent_channel_tail_burst_deletion}. We have\\
    \resizebox{\linewidth}{!}{
\begin{minipage}{\linewidth}
\begin{align*}
      \Vert B^{\leq p,\leq 2 p\cL}_{\ir}(\bx)\Vert & \leq  (968 q (n+p\cL)+1)^{2p} (2p\cL+1)(2p\cL q^{2p\cL}+1) \\
      &\leq (2p\cL+1)^2 q^{2p\cL} (968 q+1)^{2p} (n+p\cL)^{2p}\\
  &\leq q^{4p\cL}(n+p\cL)^{2p}.
\end{align*}
\end{minipage}
}\\
In the first line, $(968 q (n+p\cL)+1)^{2p}$ is derived based on Theorem~\ref{Theo:max_root_set_p_subs}; $(2p\cL+1)$ bounds the number of ways $\bs$ can be obtained from $\bx_p$ through a suffix deletion of length at most $2p\cL$; and $(2p\cL q^{2p\cL}+1)$ bounds the number of ways $\by_p$ can be obtained from $\bs$ by appending a sequence of length at most $2p\cL$. The third line is obtained by noting that $(968q+1)^{2p}(2p\cL+1)^2\leq q^{2p\cL}$ with $\cL=17$.
\end{IEEEproof}

%>>>>>>>>>>>>>>>>>>>>>>>>>>>>>>>>

\end{appendices}

\end{document}